\documentclass[11pt]{article}
\usepackage[T1]{fontenc}
\usepackage{amsmath,amssymb,amsthm,mathrsfs,amsfonts,euscript,dcolumn,color,graphicx,graphics,setspace,latexsym,lscape,subfigure,placeins,epsfig,hyperref}
\hypersetup{
    colorlinks,
    citecolor=blue,
    filecolor=black,
    linkcolor=red,
    urlcolor=magenta
}
\usepackage{euscript}
\usepackage{color}

\newcommand{\C}{\mathbb{C}}

\newcommand{\R}{\mathbb{R}}
\newcommand{\N}{\mathbb{N}}

\renewcommand{\d}{\mathrm{d}}

\newcommand{\dvol}{\mathrm{dVol}}

\newcommand{\scrh}{{\mathscr H}}
\newcommand{\scri}{{\mathscr I}}
\newcommand{\scrm}{{\mathscr M}}

\newcommand{\scrs}{{\mathscr S}}

\def\scrmhat{\hat{\scrm}}

\newtheorem{definition}{Definition}[section]
\newtheorem{theorem}{Theorem}
\newtheorem{proposition}{Proposition}[section]

\theoremstyle{remark}
\newtheorem{remark}{Remark}[section]
\begin{document}\mbox{}

\vspace{0.25in}

\begin{center}
\bf{\huge The conformal approach to asymptotic analysis}

\vspace{0.25in}

\vspace{0.1in}

{Jean-Philippe NICOLAS}\\
\small{\it{Department of Mathematics,}} \\
\small{\it{University of Brest, 6 avenue Victor Le Gorgeu,}} \\
\small{\it{29200 Brest, France.}} \\
\small{\it{Email~: Jean-Philippe.Nicolas@univ-brest.fr}}
\end{center}
\vspace{0.2in}

\begin{center}
{\bf Abstract.}
\end{center}
\noindent This essay was written as an extended version of a talk given at a conference in Strasbourg on ``Riemann, Einstein and geometry'', organized by Athanase Papadopoulos in September 2014. Its aim is to present Roger Penrose's approach to asymptotic analysis in general relativity, which is based on conformal geometric techniques, focusing on historical and recent aspects of two specialized topics~: conformal scattering and peeling.
\section{Introduction}

Hermann Minkowski, in his famous speech at the $80^\mathrm{th}$ Assembly of German Natural Scientists and Physicians in Köln in 1908, cast in a rather emphatic way the mould for what would, from then on, be the framework of relativistic mathematical physics~: ``{\it The views of space and time which I wish to lay before you have sprung from the soil of experimental physics, and therein lies their strength. They are radical. Henceforth space by itself, and time by itself, are doomed to fade away into mere shadows, and only a kind of union of the two will preserve an independent reality}.'' The last sentence is the founding principle of the geometrical description of special relativity and by extension one of the founding principles of general relativity. The new framework was to retain the essential features of Riemannian geometry, but to incorporate time into the picture and provide an indefinite metric as an intrinsic geometrical way of comparing the speed of a particle to that of light. Hermann Minkowski was not merely attempting to introduce a convenient geometrical framework for special relativity, one which was to become known as the Minkowski spacetime. He was advocating, {\it in view of physical evidence}, to give up the notion of simultaneity~; to give up the picture of the universe as a space that changes with time~; to replace it by a spacetime that is not to be understood as the succession of instants glued together, but as a global object.

When in 1917 Karl Schwarzschild discovered his famous static spherically symmetric solution of the Einstein vacuum equations, it was misunderstood as having a spherical singularity. It is precisely the change of viewpoint urged by Minkowski that allowed to replace the Schwarzschild coordinate system, tied in with a foliation by spacelike slices in the exterior region, by the Eddington-Finkelstein coordinates, based on principal null geodesics\footnote{These coordinate systems are now called Eddington-Finkelstein coordinates, as they were discovered first by Eddington in 1924 then re-discovered by Finkelstein in 1958.}, thus leading to our present understanding of the nature of the event horizon. But of course, the nature of a truly $4$-dimensional reality goes against our intuition and our experience of space and time. This is probably the reason why even though we use spacetimes every day in relativity, we are also tempted to do away with a founding principle by breaking them into a succession of spaces, what we call a $3+1$, or an $n+1$, decomposition. We are attached to such notions of simultaneity. The Cauchy problem, the related notion of global hyperbolicity, the constraints for the Einstein equations or for other overdetermined systems of equations such as Maxwell, Rarita-Schwinger, are all based on this type of decomposition. In fact, and I mention this at the risk of appearing dogmatic, even the choice of signature $-+++$, as opposed to $+---$, is prompted by similar motivations~: the reason why people choose the latter signature is usually because the metric allows to measure the proper time along causal curves (i.e. curves that are timelike or null at each point), which is in the spirit of Minkowski's geometrical approach~; the standard reason for choosing the former is that the induced metric on a spacelike slice is positive, rather than negative, definite.

The objects that our senses, our eyes in particular, allow us to observe in nature, are causal and in most cases in fact null. Of course we have access to spacelike structures but as cuts of causal objects, a topological spacelike sphere as a cut of a light-cone for example, and these cuts have a degree of arbitrariness. In a truly $4$-dimensional approach to general relativity, it seems that causal objects should be given a certain preference. As an illustration, the Cauchy problem can be replaced by the Goursat problem, the only difference being that data are now set on a null hypersurface, typically a null cone, instead of a spacelike Cauchy slice. One must be wary of a ``perversion'' of this notion whereby the initial null hypersurface is the union of two intersecting null slices meeting on a spacelike submanifold. This type of hypersurface, insofar as it is based on a spacelike structure, is not more natural than a Cauchy hypersurface. The reason why this type of problem is considered at all is because of its relative simplicity compared to the light-cone case~; the singularity of the hypersurface is very mild by comparison and the situation has advantageous similarities with the $1+1$ dimensional case. The drawback of the Goursat problem on a lightcone is that it is usually a local problem in the neighbourhood of the vertex, lightcones in generic spacetimes tending to develop caustics. Scattering theory is a global problem that can be understood as an analogue of the Goursat problem for a light-cone at infinity. The whole evolution of the field is then summarized in an operator acting between null asymptotic structures, by-passing the Cauchy problem as an unnecessary intermediate stage.

Some of us are less than others prone to thinking in $3+1$ dimensions. Roger Penrose certainly seems to have remained very faithful to Minkowski's viewpoint. His view of relativity appears to be truly $4$-dimensional. He has not systematically avoided $3+1$ decompositions and indeed has had major inputs within this approach, particularly in relation to the notion of mass, but a very significant proportion of his research is concerned with the light-cone structure (i.e. the conformal structure) of spacetime. His geometrical ideas have yielded new methods for analysis in general relativity. In this paper, I will focus on his notion of conformal compactification and how it can be used to study two types of asymptotic analytic questions~: peeling and scattering.

This paper is organized as follows. Section \ref{CC} is devoted to the description of the principles of conformal compactification, the explicit treatment of the case of Minkowski spacetime and how it provides asymptotic information on a large class of solutions of conformally invariant field equations. Section \ref{P} starts with a historical presentation of the notion of peeling, proposes a different way of looking at the question and an alternative approach to studying it, which are strongly inspired by the ideas of Roger Penrose, and finally presents the recent results in the field. Scattering, or rather a version of scattering based on conformal compactifications, is the object of section \ref{CS}~: the history of the topic is described from the founding idea by R. Penrose to the first actual construction by F.G. Friedlander~; a new approach is proposed which, by giving up some of the analytic niceties of F.G. Friedlander's results, allows to extend the construction to much more general situations~; finally the recent results following this approach are reviewed. Section \ref{Concl} contains concluding remarks.

\section{Conformal compactification} \label{CC}

\subsection{The principles of conformal compactification}

The notion of conformal compactification in general relativity was introduced by R. Penrose in a short note \cite{Pe1963} in Physical Review Letters in 1963. Its usage as a tool for studying asymptotic properties is clearly mentioned but not developed. The year after, in Les Houches, he gave a series of three lectures \cite{Pe1964} explaining the technique in details and the differences depending on the sign of the cosmological constant $\Lambda$. In 1965, specializing to the case where $\Lambda =0$, he published a long and thorough study of the asymptotic behaviour of zero rest-mass fields by means of the conformal technique \cite{Pe1965}. Another reference where a clear and detailed description of the method can be found is Spinors and Spacetime vol. 2 \cite{PeRi1984}.

There are two essential ingredients. The first is a geometrical construction~: the conformal compactification itself. It can be presented in a very general manner as follows.
\begin{itemize}
\item The ``physical'' spacetime is the spacetime on which we wish to study asymptotic properties, of test fields for example. It is a smooth, $4$-dimensional, real, Lorentzian manifold $({\cal M} , g)$.
\item The ``unphysical'', or ``compactified'', spacetime is a smooth manifold $\bar{\cal M}$ with boundary $\cal B$ and interior $\cal M$.
\item The link between the boundary and the physical spacetime is provided by a boundary defining function $\Omega$~; it is a positive function on $\cal M$, smooth on $\bar{\cal M}$, such that $\Omega \vert_{\cal B} =0$ and $\d \Omega \vert_{\cal B} \neq 0$.
\item The metric $\hat{g} := \Omega^2 g$ extends as a smooth non degenerate Lorentzian metric on $\bar{\cal M}$ (hence the name ``conformal factor'' for $\Omega$).
\end{itemize}
This conformal ``compactification''\footnote{The word compactification is a little misleading since in general the unphysical spacetime will not be compact, there will be holes in the boundary. Only in exceptional cases such as Minkowski spacetime will the rescaled spacetime be compact.} is not always possible. The property for a spacetime to admit a smooth conformal compactification can be characterized in terms of decay of the Weyl curvature at infinity. When such a compactification exists, the boundary $\cal B$ will have a structure~: different parts corresponding to different ways of going to infinity in the physical spacetime. Different parts of the boundary will play a role in relation to different types of asymptotic properties, for example timelike decay and scattering, when studied by means of the conformal method, will not involve the same parts of the conformal boundary.

The second ingredient is the equation we wish to study on the physical spacetime. It is important that it admits some rather explicit transformation law under conformal rescalings, so that we can study it on the rescaled spacetime and gain information on its behaviour in the physical spacetime. Conformally invariant equations are the natural class to consider, but not the only possible class.

\subsection{Conformal compactification of Minkowski spacetime}

Let us now present the conformal method in more details on an explicit example~: the simple case of the wave equation on flat spacetime. The contents of this section, and much more, can be found in \cite{Pe1965}.

\subsubsection{The geometrical construction}

The Minkowski metric in spherical coordinates is expressed as
\[ \eta = \d t^2 - \d r^2 - r^2 \d \omega^2 \, ,~ \d \omega^2 = \d \theta^2 +\sin^2 \theta \, \d \varphi^2 \, . \]
We choose the advanced and retarded coordinates
\begin{equation} \label{EddFinkCoord}
u= t-r \, ,~ v=t+r \, .
\end{equation}
The metric $\eta$ in terms of these coordinates takes the form
\[ \eta = \d u \d v - \frac{(v-u)^2}{4} \d \omega^2 \, , ~ v-u \geq 0 \, .\]
The we introduce new null coordinates that allow to describe the whole of Minkowski spacetime as a bounded domain~:
\begin{equation} \label{CompEddFinkCoord}
p = \arctan u \, ,~ q = \arctan v \, .
\end{equation}
The expression of the Minkowski metric in coordinates $(p,q,\omega )$ is given by
\[ \eta = (1+u^2) (1+v^2) \d p \, \d q  - \frac{(v-u)^2}{4} \d \omega^2 \, .\]
From $p$ and $q$ we can define new time and space coordinates as follows,
\begin{equation} \label{CompTimeSpaceCoord}
\begin{array}{c} {\tau = p+q = \arctan (t-r) + \arctan (t+r ) \, ,} \\ {\zeta = q-p = \arctan (t+r ) - \arctan (t-r) \, ,} \end{array}
\end{equation}
and we get
\[ \eta = \frac{(1+u^2) (1+v^2)}{4} \left( \d \tau^2 - \d \zeta^2 \right) - \frac{(v-u)^2}{4} \d \omega^2 \, .\]
We choose the conformal factor
\begin{equation} \label{FullConfFact}
\Omega = \frac{2}{\sqrt{(1+u^2)(1+v^2)}} = \frac{2}{\sqrt{(1+\tan^2 p )(1+\tan^2 q)}} = 2 \cos p \cos q  \, ,
\end{equation}
i.e. we rescale the metric by $1/r^2$ in null directions, $1/r^4$ in spacelike directions and $1/t^4$ in timelike directions. We obtain
\begin{eqnarray*}
\mathfrak{e} &:=& \Omega^2 \eta = \d \tau^2 - \d \zeta^2 - \frac{(v-u)^2}{(1+u^2)(1+v^2)} \d \omega^2 \\
&=& \d \tau^2 - \d \zeta^2 - \left( \left( \tan q - \tan p \right)  \cos p \cos q \right)^2 \d \omega^2 \\
&=& \d \tau^2 - \d \zeta^2 - \left( \sin q \cos p- \sin p \cos q \right)^2 \d \omega^2 \\
&=& \d \tau^2 - \d \zeta^2 - \left( \sin (q - p) \right)^2 \d \omega^2 \\
&=& \d \tau^2 - \d \zeta^2 - \left( \sin \zeta \right)^2 \d \omega^2 \\
&=& \d \tau^2 - \sigma^2_{S^3} \, ,
\end{eqnarray*}
where $\sigma_{S^3}^2$ is the euclidean metric on the $3$-sphere.
Minkowski spacetime is now described as the diamond
\[ \mathbb{M} = \{ |\tau| + \zeta < \pi \, ,~ \zeta \geq 0 \, ,~ \omega \in S^2 \} \, . \]
The metric $\mathfrak{e}$ is the Einstein metric~; it extends analytically to the whole Einstein cylinder $\mathfrak{E} = \R_\tau \times S^3_{\zeta, \theta , \varphi}$. The full conformal boundary of Minkowski spacetime can be defined in this framework. It is described as
\[ \partial \mathbb{M} = \{ |\tau| + \zeta = \pi \, ,~ \zeta \geq 0 \, ,\omega \in S^2 \} \, . \]
Several parts can be distinguished.
\begin{itemize}
\item Future and past null infinities~:
\begin{eqnarray*}
\scri^+ &=& \left\{ (\tau \, ,~ \zeta \, ,~ \omega ) \, ;~ \tau + \zeta = \pi \, ,~ \zeta \in ] 0,\pi [ \, ,~ \omega \in S^2 \right\} \, ,\\
\scri^- &=& \left\{ (\tau \, ,~ \zeta \, ,~ \omega ) \, ;~ \zeta -\tau = \pi \, ,~ \zeta \in ] 0,\pi [ \, ,~ \omega \in S^2 \right\} \, .
\end{eqnarray*}
\begin{proposition}
The hypersurfaces $\scri^\pm$ are smooth null hypersurfaces for $\mathfrak{e}$ (hence the terminology ``null infinities''). Their null generators are respectively the vector fields
\[ \partial_\tau - \partial_\zeta \mbox{ for } \scri^+ \mbox{ and } \partial_\tau + \partial_\zeta \mbox{ for } \scri^- \, .\]
\end{proposition}
{\bf Proof.} They are clearly smooth hypersurfaces since $\mathfrak{e}$ is analytic up to $\scri^\pm$ and does not degenerate there~: its determinent
\[ \det \left( \mathfrak{e} \right) = - \sin^4 \zeta \, . \sin^2 \theta\]
does not vanish on $\scri^\pm$ (except for the usual singularity due to spherical coordinates). Now the vector fields $\partial_\tau - \partial_\zeta$ and $\partial_\tau + \partial_\zeta$ are null and tangent respectively to $\scri^+$ and $\scri^-$. They are orthogonal to the two other generators of $\scri^\pm$~: $\partial_\theta$ and $\partial_\varphi$. They are therefore normal to $\scri^+$ and $\scri^-$ respectively. This proves the proposition. \qed
\item Future and past timelike infinities~:
\[ i^\pm = \left\{ (\tau =\pm \pi \, ,~ \zeta =0 \, ,~ \omega ) \, ;~ \omega \in S^2 \right\} \, . \]
They are smooth points for $\mathfrak{e}$ ($2$-spheres whose area is zero because they correspond to $\zeta =0$).
\item Spacelike infinity~:
\[ i^0 = \left\{ (\tau =0 \, ,~ \zeta =\pi \, ,~ \omega ) \, ;~ \omega \in S^2 \right\} \, . \]
It is also a smooth point for $\mathfrak{e}$.
\end{itemize}
Note that the Einstein spacetime $(\mathfrak{E} , \mathfrak{e})$ is static~: in the coordinates $\tau, \zeta, \theta, \varphi$, it is obvious that $\partial_\tau$ is a global timelike Killing vector field, orthogonal to the level hypersurfaces of $\tau$, which are $3$-spheres.

\begin{center}
\begin{figure}[h] \hspace{0.25cm} \begin{minipage}[c]{7cm}
\centering
\includegraphics[width=7.5cm]{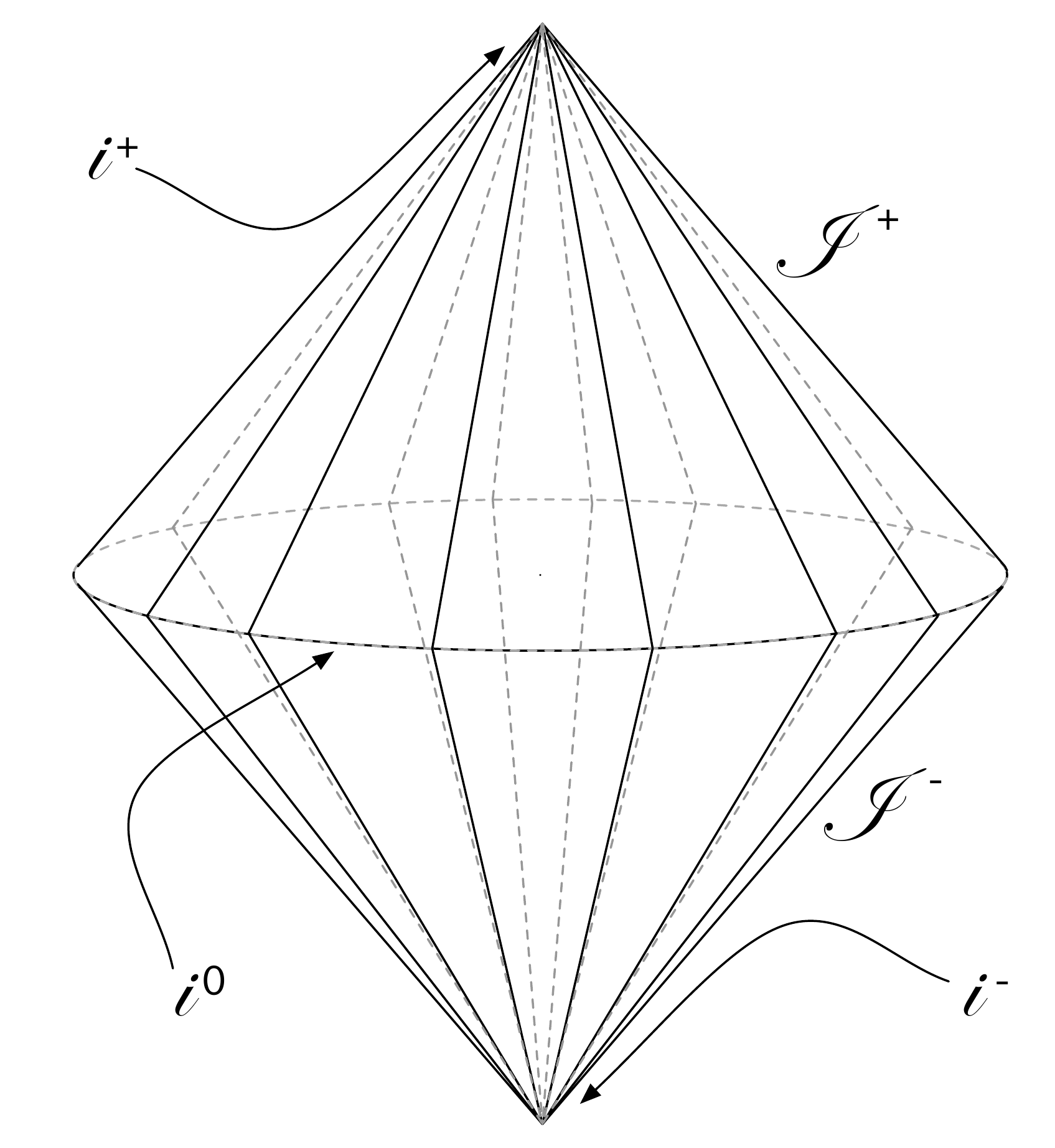}
\caption{Compactified Minkowski spacetime, $i^0$ is merely a point, just like $i^\pm$.} \label{Fig1}
\end{minipage}
\hspace{0.5cm}
\begin{minipage}[c]{7.5cm}
\centering
\includegraphics[width=1.7in]{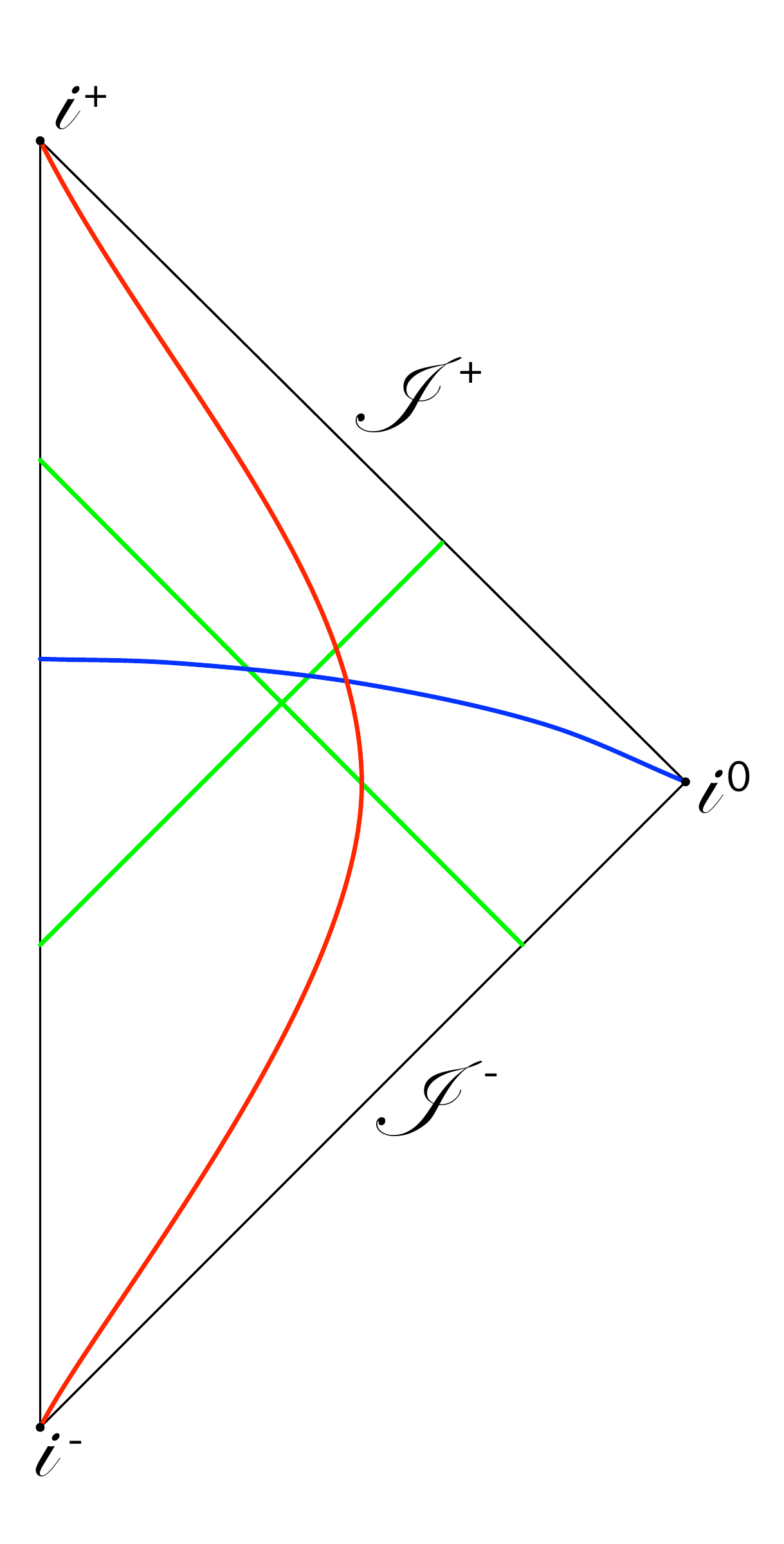}
\caption{Penrose diagram of compactified Minkowski spacetime with spacelike, timelike and null geodesics.} \label{Fig2}
\end{minipage}
\end{figure}
\end{center}

In Figure \ref{Fig1}, the conformal boundary of Minkowski spacetime is represented with its different parts. In Figure \ref{Fig2} we display the Penrose diagram of compactified Minkowski spacetime, i.e. a representation of $\overline{\mathbb{M}} = \mathbb{M} \cup \partial \mathbb{M}$ quotiented by the group of isometries inherited from the group of rotations in $\mathbb{M}$~: the spherical degrees of freedom do not appear, the advantage is that the causal structure is clearly readable on the resulting $2$-dimensional diagram.

\subsubsection{An application for a conformally invariant equation}

Let us consider a simple example of conformally invariant equation, the conformal wave equation
\begin{equation}
(\square_g + \frac{1}{6} \mathrm{Scal}_g) \phi=0 \, .
\end{equation}
Its conformal invariance can be expressed precisely as follows~: we consider a spacetime $({\cal M},g)$ and a metric $\hat{g}$ in the conformal class of $g$ with conformal factor $\Omega$, i.e. $\hat{g} = \Omega^2 g$. Then we have the equality of operators acting on scalar fields on $\cal M$
\begin{equation} \label{ConfTransWEq}
\square_g + \frac{1}{6} \mathrm{Scal}_g = \Omega^3 \left( \square_{\hat{g}} + \frac{1}{6} \mathrm{Scal}_{\hat{g}} \right) \Omega^{-1} \, ,
\end{equation}
which also entails the expression of the scalar curvature of $\hat{g}$ in terms of that of $g$
\begin{equation}
\mathrm{Scal}_{\hat{g}} = \Omega^{-2} \mathrm{Scal}_g +6 \Omega^{-3} \square_g \Omega \, .
\end{equation}
Minkowski spacetime is flat, the scalar curvature vanishes, whereas the scalar curvature of the metric $\mathfrak{e}$ is equal to $6$. Hence, a distribution $\phi \in {\cal D}' (\R^4)$ satisfies the wave equation
\begin{equation} \label{FlatWaveEq}
\partial_t^2 \phi - \Delta \phi =0 \, ,
\end{equation}
if and only if $\tilde{\phi} := \Omega^{-1} \phi$ ($\Omega$ defined by \eqref{FullConfFact}) satisfies
\begin{equation} \label{CWEEinstein}
\square_{\mathfrak{e}} \tilde{\phi} + \tilde{\phi} =0 \, ,
\end{equation}
where
\[ \square_{\mathfrak{e}} = \partial_\tau^2 - \Delta_{S^3} \, .\]
Since the Einstein cylinder is globally hyperbolic, for smooth data on $S^3$, the Cauchy problem for \eqref{CWEEinstein} has a unique smooth solution on the whole of $\R_\tau \times S^3$ (see J. Leray \cite{Le1953}). Let us consider data $\phi_0 \, , ~\phi_1$ at $t=0$ for the Cauchy problem for \eqref{FlatWaveEq}, i.e.
\begin{equation} \label{PhysData}
\phi_0 = \phi \vert_{t=0} \, ,~ \phi_1 = \partial_t \phi \vert_{t=0} \, .
\end{equation}
From these, we can easily calculate the corresponding data at $\tau =0$ for $\tilde\phi$, i.e.
\begin{equation} \label{RescData}
\tilde\phi_0 = \tilde\phi \vert_{\tau=0} \, ,~ \tilde\phi_1 = \partial_\tau \tilde\phi \vert_{\tau=0} \, ,
\end{equation}
using the fact that on $\mathbb{M}$, $t=0$ is equivalent to $\tau =0$. First it is immediate that
\[ \tilde{\phi}_0 = \left( \Omega \vert_{t=0} \right)^{-1} \phi_0 = \frac{1+r^2}{2} \phi _0 \, .\]
Concerning the other part of the data,
\[ \partial_t \phi = \left( \partial_t \Omega \right) \tilde{\phi} + \Omega \frac{\partial \tau}{\partial t} \partial_\tau \tilde{\phi} + \Omega \frac{\partial \zeta}{\partial t} \partial_\zeta \tilde{\phi} \]
and
\[ \frac{\partial \Omega}{\partial t} \vert_{t=0} = 0 \, ,~ \frac{\partial \tau}{\partial t} \vert_{t=0} = \frac{2}{1+r^2} \, ,~ \frac{\partial \zeta}{\partial t} \vert_{t=0} = 0 \, ,\]
which gives
\[ \partial_t \phi_{|_{t=0}} = \frac{4}{(1+r^2)^2} \partial_\tau \tilde{\phi}_{|_{\tau =0}} = \frac{4}{(1+r^2)^2} \tilde{\phi}_1 \, . \]
Hence the relation between the data \eqref{RescData} for the rescaled field and \eqref{PhysData} for the physical field~:
\begin{equation} \label{RelData}
\tilde{\phi}_0 = \frac{1+r^2}{2} \phi_0 \, ,~ \tilde{\phi}_1 = \frac{(1+r^2)^2}{4} \phi_1 \, .
\end{equation}
Let us make the assumption that $\tilde{\phi}_0$ and $\tilde{\phi}_1$, which are naturally defined only on the $3$-sphere at $\tau=0$ with the point $i^0$ removed, extend as smooth functions on $S^3$. Then the rescaled solution $\tilde{\phi} = \Omega^{-1} \phi$ extends as a smooth function on $\overline{\mathbb{M}}$ and by simple explicit calculations, we can infer precise pointwise decay rates of the unrescaled field in all causal directions. All we need is to use the continuity of $\tilde{\phi}$ at the boundary of $\overline{\mathbb{M}}$ and the behaviour of the conformal factor $\Omega$ along null and timelike geodesics (easily obtained from \eqref{FullConfFact}).
\begin{enumerate}
\item {\bf Decay along null directions.} There exist smooth functions $\tilde{\phi}^\pm \in {\cal C}^\infty (\R \times S^2 )$ such that for all $u,v,\omega$,
\begin{gather}
\lim_{r\rightarrow +\infty} r \phi (t= r + u , r, \omega ) = \frac{1}{\sqrt{1+u^2}} \, \tilde{\phi}^+ (u,\omega )\, , \label{FRF}\\
\lim_{r\rightarrow +\infty} r \phi (t= -r + v , r, \omega ) = \frac{1}{\sqrt{1+v^2}} \, \tilde{\phi}^- (v,\omega )\, . \label{PRF}
\end{gather}
The functions $\tilde{\phi}^\pm$ are simply the restrictions of $\tilde{\phi}$ on $\scri^\pm$~; the two limits above are referred to as the future and past asymptotic profiles, or radiation fields, of $\phi$.
\item {\bf Decay along timelike directions.} There exist two constants $C^\pm$ such that for all $r,\omega$,
\[ \lim_{t\rightarrow \pm\infty} t^2 \phi (t, r, \omega ) = 2C^\pm \, . \]
These constants are $C^\pm = \tilde{\phi} (i^\pm )$ (recall that $i^\pm$ are points on the Einstein cylinder, not $2$-spheres).
\end{enumerate}
In other words, the physical solution $\phi$ decays like $1/r$ along radial null geodesics and like $1/t^2$ along the integral lines of $\partial_t$. These decay rates are valid for solutions $\phi$ of the wave equation on Minkowski spacetime such that $\tilde{\phi} = \Omega^{-1} \phi$ extends as a smooth function on $\mathfrak{E}$. Implicit in this hypothesis are some requirements on the fall-off of initial data for $\phi$. Using \eqref{RelData}, the smoothness of $\tilde{\phi}_0$ and $\tilde{\phi}_1$ on $S^3$ entails that there exist two constants $C_0 , C_1$ such that for all $\omega$,
\begin{gather*}
\lim_{r\rightarrow +\infty} r^2 \phi (0, r , \omega ) = 2 C_0 \, ,\\
\lim_{r\rightarrow +\infty} r^4 \partial_t \phi (0, r , \omega ) = 4 C_1 \, ,
\end{gather*}
the constants $C_0$ and $C_1$ being the respective values of $\tilde{\phi}_0$ and $\tilde{\phi}_1$ at $i^0$ (which, like $i^\pm$ is a point on $\mathfrak{E}$).

The crucial observation which allowed us to derive the decay rates above is that the information on the pointwise decay of $\phi$ at infinity is equivalent to the continuity of the rescaled field at the conformal boundary. It is possible to go further. In the next two sections, we present two refinements of this first use of conformal compactification for asymptotic analysis.

\section{Peeling} \label{P}

The peeling, or peeling-off of principal null directions, is a generic asymptotic behaviour discovered by R. Sachs for spin $1$ and $2$ fields in the flat case \cite{Sa61} and in the asymptotically flat case \cite{Sa62}. A zero-rest-mass field of spin $s$ is described as a symmetric spinor of rank $2s$. Such an object possesses $2s$ principal null directions at each point, which are analogous to the roots of a polynomial of degree $2s$. An outgoing zero rest-mass field of spin $s$, along a null geodesic going out to infinity, can be expressed as an expansion in powers of $1/r$. This expansion is such that the part of the field falling-off like $r^{-k}$, $1\leq k \leq 2s$, has $2s-k$ of its principal null directions aligned along the null geodesic. The notion was explored further by E.T. Newman and R. Penrose \cite{NP} and by R. Penrose \cite{Pe1963,Pe1965}, using the spin-coefficient formalism (now referred to as the Newman-Penrose formalism) and the conformal method. In \cite{Pe1965}, the conformal method is used in order to show that the peeling-off of principal null directions is equivalent to a very simple property~: the boundedness of the rescaled field at null infinity. The question I will focus my attention on is that of the genericity of the peeling behaviour. It is a delicate question which remained controversial for some years. The reason for this controversy was the logarithmic divergence observed when comparing the asymptotic structures of Minkowski and Schwarzschild spacetimes. This is expected to be generic since a physically relevent asymptotically flat spacetime ought to be a short-range perturbation of a Schwarzschild spacetime. The question was however answered in the affirmative in \cite{MaNi2009,MaNi2012} by treating the Schwarzschild case. The key idea was to reformulate the peeling property in terms of energy estimates instead of pointwise behaviour along outgoing null geodesics. This section is a description of these results and of the path that led to them.

\subsection{A new approach to the peeling}

As we saw above, the peeling in its original form is equivalent to the boundedness of the rescaled field at null infinity. It is naturally tempting to define a notion of peeling of higher order, corresponding to higher degrees of regularity at the conformal boundary. This is exactly what Roger Penrose did when he defined $k$-asymptotically simple spacetimes as a generic model for asymptotic flatness (see \cite{PeRi1984} Vol. 2). A
4-dimensional, globally hyperbolic, Lorentzian space-time $\left( \scrm, g\right)$, $\scrm \simeq \R^4$, is called $k$ asymptotically simple if there exists a globally hyperbolic ${\cal C}^{k+1}$ Lorentzian manifold $(\scrmhat , \hat{g} )$ with boundary $\scri$ and a scalar field  $\Omega$ on $\scrmhat$ such that~:
\begin{description}
\item[(i)] $\scrm$ is the interior of $\scrmhat$~;
\item[(ii)] $\hat{g}_{ab}= \Omega^2 g_{ab}$ on $\scrm$~;
\item[(iii)] $\Omega$ and $\hat{g}$ are ${\cal C}^k$ on $\scrmhat$~;
\item [(iv)] $\Omega >0$ on $\scrm$~; $\Omega =0$ and $\d \Omega \neq 0$ on $\scri$~;
\item[(v)] every null geodesic in $\scrm$ acquires a future endpoint on $\scri^+$ and a past endpoint on $\scri^-$.
\end{description}
However, when defining peeling of higher order for zero rest-mass fields, it is not convenient to use ${\cal C}^k$ spaces to characterize their regularity at $\scri$. The reason is that the Cauchy problem for hyperbolic equations is not well-posed in ${\cal C}^k$ spaces.

In a large portion of the litterature concerned with the peeling and particularly its genericity, what seems to have been lacking is a precise definition of what one is really trying to prove or disprove. For example, saying that on a given asymptotically flat spacetime there is no peeling would not make much sense, because, unless the asymptotic flatness is too week to even define $\scri$ as a conformal boundary, the chances are that there will always be conformally rescaled fields that extend continuously at $\scri$~; smooth compactly supported initial data would usually guarantee such a behaviour. Also finding examples of data for which the rescaled field does not extend continuously at $\scri$ is no proof that the asymptotic structure of the spacetime is radically different from that of Minkowski spacetime. Indeed, in the flat case, if we take for the wave equation smooth initial data that are, say, exponentially increasing at spacelike infinity, the rescaled field will not even be bounded at $\scri$. An important information however, would be, on a given asymptotically flat spacetime, to have data such that the associated solution does not peel, but whose regularity and decay at spacelike infinity is enough to entail peeling in Minkowski spacetime. As we shall see below, such a situation is however very unlikely, certainly it is not possible on the Schwarzschild metric for the wave equation, Dirac or Maxwell fields. Here is a precise way of addressing the question of the genericity of the peeling, or rather its higher order version, in the form of a scheme in two steps~:
\begin{description}
\item[Step 1.] Characterize, on a given asymptotically flat spacetime, the class of data that ensures a given regularity of the rescaled field at $\scri$.
\item[Step 2.] Compare such classes between different spacetimes, in particular, compare them with the corresponding classes in the case of Minkowski spacetime.
\end{description}
It remains to decide how to measure the regularity of the rescaled field at $\scri$ and how to proceed to obtain the optimal classes of data ensuring such regularity. I shall adopt four main guiding principles to do so.
\begin{enumerate}
\item {\bf Work on the compactified spacetime~:}  formulate the peeling in terms of regularity at $\scri$, not in terms of an asymptotic expansion along null directions. The reason for this is that pointwise regularity at $\scri$ can be precisely controlled, without loss of information, in terms of regularity and decay of initial data. This is less true of the asymptotic behaviour of fields described in terms of finite asymptotic expansions, particularly when we are after a one to one correspondance between an order of expansion and a class of data.
\item {\bf Work in a neighbourhood of spacelike infinity.} Once the regularity is established at $\scri$ near spacelike infinity, it can easily be inferred further up $\scri$ provided the solution is smooth enough in the bulk. What we wish to avoid is singularities creeping up $\scri$ due to insufficient decay assumptions on the data.
\item {\bf Avoid the use of ${\cal C}^k$ spaces.} The first natural idea is to consider that a field peels at order $k$ if the rescaled field is ${\cal C}^k$ at $\scri$. This is a perfectly valid definition but, because of the ill-posedness of the Cauchy problem in ${\cal C}^k$ spaces, it makes it difficult to characterize this behaviour by a class of initial data. Instead, I will characterize the order of peeling using Sobolev spaces whose norms are energy fluxes and for which the Cauchy problem is naturally well-posed. The optimal class of data for which fields peel at a given order can then be studied using geometric energy estimates.
\item {\bf Avoid Sobolev embeddings.} A common approach in the study of decay of fields is to use integrated energy estimates and to turn them into pointwize estimates via Sobolev embeddings~: by means of geometric energy estimates, one controls weighted Sobolev norms in the bulk, then Sobolev inequalities, by providing an embedding of the relevent weighted Sobolev space into a weighted ${\cal C}^k$ space, give locally uniform pointwize decay rates. The problem is that Sobolev embeddings lose derivatives. Besides, they are valid for an open set of regularities so estimates in fact always lose a little more derivatives than necessary. The method is therefore not adapted to finding the optimal class of data ensuring a certain regularity at the conformal boundary. The solution is to define the peeling of a given order as a Sobolev regularity of the rescaled field at $\scri$. It is precise, does not involve conversion with loss between Sobolev and ${\cal C}^k$ regularities and allows optimal control in terms of the regularity of the data by means of energy estimates. The kind of estimates we will use here are not integrated estimates, but estimates between the energy fluxes on $\scri$ and on a Cauchy hypersurface. Their optimality will be ensured by imposing that they be valid both ways. This  essentially means that we prove the equivalence of the energies on $\scri$ and the Cauchy hypersurface. The fact that we work in a neighbourhood of spacelike infinity and not on the whole spacetime merely requires to have an added term in the estimates, corresponding to the flux of energy leaving the neighbourhood of $i^0$.
\end{enumerate}

\subsection{A first natural framework in the flat case} \label{FullCompact}

For a first approach, we use the complete regular conformal compactification on Minkowski spacetime to perform global energy estimates. These estimates provide us with a definition and characterization of the peeling at any order for the wave equation.

Let us consider the stress energy tensor for equation \eqref{CWEEinstein}
\begin{equation} \label{SETensorEinsWE}
\tilde{T}_{ab} = \partial_a \tilde{\phi} \partial_b \tilde{\phi} - \frac12 \mathfrak{e}_{ab} \mathfrak{e}^{cd} \partial_c \tilde{\phi} \partial_d \tilde{\phi} +\frac{1}{2} \tilde{\phi}^2 \mathfrak{e}_{ab} \, .
\end{equation}
It is symmetric and divergence-free when $\tilde{\phi}$ is a solution of \eqref{CWEEinstein} since
\[ \tilde{\nabla}^a \tilde{T}_{ab} = ( \square_{\mathfrak{e}} \tilde{\phi} + \tilde{\phi} ) \partial_b \tilde{\phi} \, ,\]
where $\tilde{\nabla}$ denotes the Levi-Civita connection associated with the Einstein metric $\mathfrak{e}$. Hence, contracting $\tilde{T}_{ab}$ with the Killing vector field $K = \partial_\tau$, we have the conservation law
\begin{equation} \label{EinsteinConsLaw}
\nabla^a \left( K^b \tilde{T}_{ab} \right) = 0 \, .
\end{equation}
The vector field $J^a := K^b \tilde{T}_b^a$ is the energy current that we shall use for the estimates. On a given oriented piecewise ${\cal C}^1$ hypersurface $S$, the flux of $J$ is given by
\[ {\cal E}_{K,S} (\tilde{\phi} ) = \int_{S} J_a n^a ( l \lrcorner \dvol ) \, ,\]
where $l^a$ is a vector field transverse to $S$ compatible with the orientation of $S$ and $n^a$ a normal vector field to $S$ such that $l_a n^a =1$.

For instance, denoting $X_\tau = \{ \tau \} \times S^3$ the level hypersurfaces of the function $\tau$,
\begin{equation} \label{NormS3}
{\cal E}_{K,X_\tau} (\tilde{\phi} ) = \frac{1}{2} \int_{X_\tau} \left( (\partial_\tau \tilde{\phi} )^2 + \left| \nabla_{S^3} \tilde{\phi} \right|^2 + \tilde{\phi}^2 \right) \d \mu_{S^3} = \frac12 \left( \Vert \tilde{\phi} \Vert_{H^1 (X_\tau )}^2 + \Vert \partial_\tau \tilde{\phi} \Vert_{L^2 (X_\tau )}^2 \right) \, .
\end{equation}
Also, parametrizing $\scri^+$ as $\tau = \pi - \zeta$,
\begin{eqnarray}
{\cal E}_{K,\scri^+} (\tilde{\phi} ) &=& \frac{1}{\sqrt{2}} \int_{\scri^+} \left( -2 \partial_\tau \tilde{\phi} \, \partial_\zeta \tilde{\phi} + (\partial_\tau \tilde{\phi} )^2 + \left| \nabla_{S^3} \tilde{\phi} \right|^2 + \tilde{\phi}^2 \right) \d \mu_{S^3} \nonumber \\
&=& \frac{1}{\sqrt{2}} \int_{\scri^+} \left( \left| \partial_\tau \tilde{\phi} - \partial_\zeta \tilde{\phi} \right|^2 + \frac{1}{\sin^2 \zeta} \left| \nabla_{S^2} \tilde{\phi} \right|^2 + \tilde{\phi}^2 \right) \d \mu_{S^3} \, . \label{NormScri}
\end{eqnarray}
This is a natural $H^1$ norm of $\tilde{\phi}$ on $\scri^+$, involving only the tangential derivatives of $\tilde{\phi}$ along $\scri^+$.

Now consider a smooth solution $\tilde{\phi}$ of (\ref{CWEEinstein}) on $\mathfrak{E}$. The conservation law (\ref{EinsteinConsLaw}) tells us that the flux of $J$ across the closed hypersurface made of the union of $X_0$ and $\scri^+$ is zero. Hence,
\begin{equation} \label{EnEqEin}
{\cal E}_{K,\scri^+} (\tilde{\phi} ) = {\cal E}_{K,X_0} (\tilde{\phi} ) \, .
\end{equation}
Moreover, since $\partial_\tau$ is a Killing vector, for any $k\in \N$, $\partial_\tau^k \tilde{\phi}$ satisfies equation (\ref{CWEEinstein}), whence
\[ {\cal E}_{K,\scri^+} (\partial_\tau^k \tilde{\phi} ) = {\cal E}_{K,X_0} (\partial_\tau^k \tilde{\phi} ) \, . \]
Using equation (\ref{CWEEinstein}), for $k= 2p$, $p\in \N$, we have
\begin{eqnarray}
2 {\cal E}_{K,X_0} (\partial_\tau^k \tilde{\phi} ) &=& \| \partial_\tau^{2p} \tilde{\phi} \|_{H^1 (X_0 )}^2 + \| \partial_\tau^{2p+1} \tilde{\phi} \|_{L^2 (X_0 )}^2 \nonumber \\
&=&  \| (1-\Delta_{S^3} )^{p} \tilde{\phi} \|_{H^1 (X_0)}^2 + \| (1-\Delta_{S^3} )^{p} \partial_\tau \tilde{\phi} \|_{L^2 (X_0)}^2 \nonumber \\
&\simeq & \| \tilde{\phi} \|_{H^{2p+1} (X_0)}^2 + \| \partial_\tau \tilde{\phi} \|_{H^{2p} (X_0)}^2 \, , \label{EinsteinEnEqEven}
\end{eqnarray}
and for $k=2p+1$, $p \in \N$,
\begin{eqnarray}
2 {\cal E}_{K,X_0} (\partial_\tau^k \tilde{\phi} ) &=& \| \partial_\tau^{2p+1} \tilde{\phi} \|_{H^1 (X_0 )}^2 + \| \partial_\tau^{2p+2} \tilde{\phi} \|_{L^2 (X_0 )}^2 \nonumber \\
&=& \| (1-\Delta_{S^3} )^{p} \partial_\tau \tilde{\phi} \|_{H^1 (X_0)}^2 + \| (1-\Delta_{S^3} )^{p+1} \tilde{\phi} \|_{L^2 (X_0)}^2 \nonumber \\
&\simeq & \| \tilde{\phi} \|_{H^{2p+2} (X_0)}^2 + \| \partial_\tau \tilde{\phi} \|_{H^{2p+1} (X_0)}^2\, . \label{EinsteinEnEqOdd}
\end{eqnarray}
Hence, we have for each $k \in \N$~:
\[ \| \tilde{\phi} \|^2_{H^{k+1} (X_0 )} + \| \partial_\tau \tilde{\phi} \|^2_{H^{k} (X_0 )} \simeq  {\cal E}_{K,X_0} (\partial_\tau^k \tilde{\phi} ) = {\cal E}_{K,\scri^+} (\partial_\tau^k \tilde{\phi} ) \simeq \| \partial^k_\tau \tilde{\phi} \|^2_{H^{1} (\scri^+ )} \]
and using the fact that the $H^k$ norm controls all the lower Sobolev norms, this gives us the apparently stronger equivalence

\begin{equation} \label{EinsteinEnEq}
\| \tilde{\phi} \|^2_{H^{k+1} (X_0 )} + \| \partial_\tau \tilde{\phi} \|^2_{H^{k} (X_0 )} \simeq  \sum_{p=0}^k \| \partial^p_\tau \tilde{\phi} \|^2_{H^{1} (\scri^+ )} \, .
\end{equation}
\begin{remark}
This equivalence should not in principle be understood as providing a solution to a Goursat problem on $\scri^+$. Indeed, in Lars H\"ormander's paper on the Goursat problem for the wave equation \cite{Ho1990}, it is made very clear that such an equivalence only provides us with a trace operator on $\scri^+$ that is a partial isometry, it is then necessary to prove the surjectivity of this operator in order to solve the Goursat problem. However, we know from the same paper that the Goursat problem for equation (\ref{CWEEinstein}) with data $\tilde{\phi}_{|_{\scri^+}} \in H^1 (\scri^+)$ is well posed and gives rise to solutions $\tilde{\phi} \in {\cal C}^0 (\R_\tau \, ;~H^1 (S^3 )) \cap {\cal C}^1 (\R_\tau \, ;~L^2 (S^3 ))$. Hence (\ref{EinsteinEnEq}) indeed provides us with a regularity result for the Goursat problem~: data on $\scri^+$ for which the norm on the right-hand side is finite give rise to solutions that are in ${\cal C}^l (\R_\tau \, ;~H^{k+1-l} (S^3 ))$ for all $0\leq l \leq k+1$. This is however stronger than the information we are interested in. We simply extract from (\ref{EinsteinEnEq}) the fact that for smooth solutions, the control of the transverse regularity on $\scri^+$ described by ${\cal E}_{\scri^+} (\partial_\tau^p \tilde{\phi} )$, $0\leq p \leq k$, is equivalent to that of the $H^{k+1}$ norm of the restriction of $\tilde{\phi}$ to $X_0$ and the $H^k$ norm of the restriction of $\partial_\tau \tilde{\phi}$ to $X_0$. By a standard density argument, this shows that if we wish to guarantee, by means of some control on the initial data, that the restriction to $\scri^+$ of $\partial_\tau^p \tilde{\phi}$, $0\leq p \leq k$, is in $H^1 (\scri^+)$, the optimal condition to impose is that $\tilde{\phi}_0 \in H^{k+1} (X_0)$ and $\tilde{\phi}_1 \in H^k (X_0)$. This is our first definition of a peeling of order $k$ and its characterization by a function space of initial data.
\end{remark}

\begin{definition} \label{DefPeelMinkEnEst}
A solution $\phi$ of (\ref{FlatWaveEq}) is said to peel at
order $k \in \N$ if the traces on $\scri^+$ of $\partial_\tau^p \tilde{\phi}$, $0\leq p \leq k$,
are in $H^{1} (\scri^+)$. The optimal function space of initial data $ (\tilde{\phi}_0 \, ,~ \tilde{\phi}_1 )$ giving
rise to solutions that peel at order $k$ is $H^{k+1} (S^3) \times H^k (S^3)$.
\end{definition}
Expressing the space of data in terms of the physical field $\phi$ using \eqref{RelData} gives us the exact function space of physical data giving rise to solutions of (\ref{FlatWaveEq}) that peel at order $k$.

\subsection{The ``correct'' version in the flat case}\label{MinimalCompact}

The previous construction is very natural but its drawback is that very few spacetimes admit such a complete and regular compactification. As a consequence, we may have a valid definition of the peeling at any order in the flat case, but we will not be able to compare with other asymptotically flat spacetimes, whose natural compactifications are associated with much weaker conformal factors. On the Schwarzschild spacetime for instance, we cannot hope to compactify the exterior of the black hole in as complete a manner as Minkowski spacetime. Timelike and spacelike infinities will be singularities of the conformal structure. The natural compactification is associated to $\Omega = 1/r$, which also has the pleasant property that we have the same symmetries before and after rescaling. Since the conformal factors $1/r$ and \eqref{FullConfFact} are not uniformly equivalent at infinity in the flat case, it would not make sense to compare the peeling on the Schwarzschild metric constructed with $1/r$, to the version above on Minkowski spacetime constructed with \eqref{FullConfFact}. We must therefore redefine the peeling in the flat case using $1/r$ as a conformal factor. Then we can perform similar constructions on the Schwarzschild metric and compare the results. Besides, on most asymptotically flat spacetime, it will be possible to define an analogous conformal factor that will allow to construct $\scri$. This choice of conformal factor will therefore make comparison with Minkowski spacetime relatively easy for a large class of asymptotically flat geometries.

We perform the construction on $\mathbb{M}$ purely for future null infinity. The analogous construction for $\scri^-$ is obtained by a straightforward modification. We write the Minkowski metric in terms of variables $u=t-r$, $r$, and $\omega$~:
\[ \eta = \d u^2 + 2 \d u \d r - r^2 \d \omega^2 \, .\]
Then putting $R = 1/r$, we get
\begin{equation} \label{MinkMetPartResc}
\hat{\eta} := R^2 \eta = R^2 \d u^2 -2 \d u \d R - \d \omega^2 \, .
\end{equation}
This rescaled metric extends analytically on $\R_u \times [0,+\infty [_R \times S^2_\omega$, which is Minkowski spacetime minus the $r=0$ coordinate line, with the added boundary $\R_u \times \{ 0 \}_R \times S^2_\omega$. This boundary can easily be seen to be $\scri^+$ since for $u$ and $\omega$ constant, we move on an outgoing radial null geodesic and the boundary is reached as $r \rightarrow +\infty$ along such lines. In the new coordinates $u,R,\omega$, the vector field $\partial_u$ is the timelike Killing vector field for $\eta$ that used to be $\partial_t$ in the $t,r,\omega$ coordinates. It turns out that $\partial_u$ is still Killing for the rescaled metric $\hat{\eta}$. The scalar curvature of $\hat{\eta}$ vanishes
\begin{equation} \label{ScalHatEta}
\frac{1}{6} \mathrm{Scal}_{\hat{\eta}} = R^{-3} \square_\eta R = 0 \, ,
\end{equation}
so $\phi \in {\cal D}' (\mathbb{M} )$ satisfies (\ref{FlatWaveEq}) if and only if $\hat{\phi} := R^{-1} \phi = r \phi$ is a solution of
\begin{equation} \label{FutureRescWaveEq}
\square_{\hat{\eta}} \hat{\phi} = 0 \, ,~ \square_{\hat{\eta}} = -2 \partial_u \partial_R - \partial_R R^2 \partial_R - \Delta_{S^2} \, .
\end{equation}
We work on a neighbourhood of $i^0$ of the following form
\[ \Omega_{{u_0}} = \left\{ (u,R,\omega )\, ;~ u \leq u_0 \, ,~ R \geq 0 \, ,~ t\geq 0 \right\} \, ,\]
for $u_0 << -1$. The boundary of $\Omega_{{u_0}}$ is made of three parts~:
\begin{eqnarray*}
\scri^+_{u_0} &=& \scri^+ \cap \Omega_{{u_0}} \, ,\\
\Sigma_{0}^{u_0} &=& \Sigma_0 \cap \Omega_{{u_0}} \, ,\\
{\cal S}_{u_0} &=& \{ u=u_0 \} \cap \Omega_{{u_0}} \, .
\end{eqnarray*}
The usual stress-energy tensor for the wave equation
\begin{equation} \label{SETMinkHat}
\hat{T}_{ab}=\partial_a \hat\phi \, \partial_b \hat\phi - \frac12 \hat{\eta}_{ab} \hat{\eta}^{cd} \partial_c \hat\phi \, \partial_d \hat\phi
\end{equation}
is symmetric and divergence-free for solutions of \eqref{FutureRescWaveEq} since
\[ \hat{\nabla}_a \hat{T}_{ab} = (\square_{\hat{\eta}} \hat{\phi} ) \partial_b \hat{\phi} \, ,\]
where $\hat{\nabla}$ denotes the Levi-Civita connection associated with the metric $\hat{\eta}$.

We need to find an analogue of $\partial_\tau$ on the Einstein cylinder with which to define an energy current~: a vector field that extends smoothly at the conformal boundary as a transverse vector field to $\scri$ and that is as close as possible to being Killing. We could simply re-express $\partial_\tau$ in terms of the coordinates $u,R,\omega$ but its flow would not be an isometry. It turns out that a modification of $\partial_\tau$ by a multiple of $\partial_t$ is a Killing vector field for $\hat\eta$ and is transversal to $\scri$~: it is the Morawetz vector field introduced by Kaithleen Morawetz in 1961 \cite{Mo1961} in order to prove decay estimates for the wave equation on Minkowski spacetime. It is defined in terms of variables $u$ and $v$ as
\[ T = u^2 \partial_u + v^2 \partial_v, \mbox{ i.e. } T= (t^2+r^2) \partial_t +2tr \partial_r \, .\]
and is therefore timelike everywhere on $\mathbb{M}$ except on the lightcone of the origin, where it is null. It is future-oriented on $\mathbb{M}$, except at the origin where it vanishes. In terms of the variables $u$ and $R$, it reads
\begin{equation} \label{MorawetzVF}
T = u^2 \partial_u - 2 (1+uR) \partial_R \, .
\end{equation}
It extends smoothly at $\scri^+$ where it takes the expression $u^2 \partial_u -2\partial_R$ which is the sum of two future-oriented null vector fields, one tangent to $\scri$ and the other transverse. Moreover, $T$ satisfies the Killing equation for $\hat{\eta}$~:
\[ \hat\nabla^{(a} T^{b)} = 0 \, ,\]
so we immediately get the following energy identity, for a smooth solution $\phi$ of \eqref{FlatWaveEq} with compactly supported initial data,
\begin{equation} \label{BasicIdentityMink}
{\cal E}_{T, \scri^+_{u_0}} (\hat{\phi})+ {\cal E}_{T,{\cal S}_{u_0}} (\hat{\phi}) = {\cal E}_{T,\Sigma_{0}^{u_0}} (\hat{\phi}) \, ,
\end{equation}
where
\begin{eqnarray}
{\cal E}_{T,{\Sigma}_{0}^{u_0}} (\hat\phi ) &=& \int_{{\Sigma}_{0}^{u_0}} \left( u^2 ( \partial_u \hat\phi )^2 + R^2 u^2 \partial_u \hat\phi \partial_R \hat\phi \right. \nonumber \\
&& \left. + R^2 \left( \frac{(2+uR)^2}{2}  - (1+uR) \right) (\partial_R \hat\phi )^2 + \left( \frac{u^2 R^2}{2} + 1+uR \right) \left\vert \nabla_{S^2} \hat\phi \right\vert^2 \right) \d u \d^2 \omega \nonumber \\
& \simeq & \int_{{\Sigma}_{0}^{u_0}} \left( u^2(\partial_u \hat\phi )^2 + R^2 (\partial_R \hat\phi )^2+ \left| \nabla_{S^2} \hat\phi \right|^2 \right)  \d u\d^2\omega \, , \label{EnergySigma0} \\
{\cal E}_{T, \scri^+_{u_0}} (\hat{\phi}) &=& \int_{\scri^+_{u_0}} \left( u^2(\partial_u \hat\phi )^2 + \vert\nabla_{S^2} \hat\phi \vert^2 \right) \d u \d^2 \omega \, , \label{EnergyScri+} \\
{\cal E}_{T,{\cal S}_u} (\phi ) &=& \int_{{\cal S}_u} \frac12\left((2+uR)^2 (\partial_R \hat\phi )^2+u^2 \vert\nabla_{S^2}\hat\phi \vert^2 \right) \d R \d ^2 \omega \, . \label{EnergySu}
\end{eqnarray}
The next thing to work out is how to obtain higher order estimates from \eqref{BasicIdentityMink}. We could simply commute $T$ into the equation~; since it is Killing, we would immediately obtain identitites similar to \eqref{BasicIdentityMink} for $T^k \hat{\phi}$. Although this would again be a perfectly valid definition, it would not extend naturally to other spacetimes such as Schwarzschild, because in these spacetimes, the Killing form of $T$ induces terms that are delicate to handle in the higher order estimates (see \cite{MaNi2009} for details). Another vector field that we can use is $\partial_R$. It is a null vector field, as can be seen from \eqref{MinkMetPartResc}, it is transverse to $\scri^+$ and has the following expression in terms of the variables $t,r$ (which we shall use later on)~:
\begin{equation} \label{DRdtdr}
\partial_R = -r^2 ( \partial_t + \partial_r ) \, .
\end{equation}
The vector field $\partial_R$ is not Killing for $\hat\eta$ but it is nontheless easy to control the error terms in the higher order estimates, since they merely involve polynomials in $R$. Moreover, this vector field will turn out to be just as easy to use in the Schwarzschild framework, for the same reason. Using Gronwall's inequality, we obtain the following estimates both ways~: for each $k \in \N$, there exists a positive constant $C_k$ such that, for any smooth solution $\hat\phi$ of \eqref{FutureRescWaveEq} with compactly supported initial data,
\begin{eqnarray}
{\cal E}_{T,{\scri^+_{u_0}}} (\partial_R^k \hat\phi ) &\leq & C_k \sum_{p=0}^k {\cal E}_{T,{\Sigma}_{0}^{u_0}} (\partial_R^p \hat\phi ) \, , \label{MinkHOEst1}\\
{\cal E}_{T,{\Sigma}_{0}^{u_0}} (\partial_R^k \hat\phi ) &\leq & C_k \sum_{p=0}^k \left( 
{\cal E}_{T,\scri_{u_0}^+} (\partial_R^p \hat\phi ) + {\cal E}_{T,{\cal S}_{u_0}} (\partial_R^p \hat\phi ) \right) \, . \label{MinkHOEst2}
\end{eqnarray}
Using the spherical symmetry, we can also add angular derivatives in the estimates above. These higher order estimates give us a definition of the peeling at any order on Minkowski spacetime, that is different from the one above, but that extends to other asymptotically flat spacetimes.
\begin{definition}
We say that a solution $\hat\phi$ of \eqref{FutureRescWaveEq} peels at order $k\in \N$ if for all polynomial $P$ in $\partial_R$ and $\nabla_{S^2}$ of order lower than or equal to $k$, we have ${\cal E}_{T,\scri^+_{u_0} } (P \hat\phi ) <+\infty$. This means than for all $p \in \{ 0,1,...,k\}$ we have for all $q \in \{ 0,1,...,p \}$, ${\cal E}_{T,\scri^+_{u_0} } (\partial_R^q \nabla_{S^2}^{p-q} \hat\phi ) <+\infty$.
\end{definition}
In view of estimates \eqref{MinkHOEst1}, \eqref{MinkHOEst2} the condition on initial data that guarantees peeling at order $k$ is therefore that
\[ \forall p \in \{ 0,1,...,k\} \, ,~ \forall q \in \{ 0,1,...,p \} \, , ~ {\cal E}_{T,{\Sigma}_{0}^{u_0} } (\partial_R^q \nabla_{S^2}^{p-q} \hat\phi ) <+\infty \, .\]
This can easily be re-expressed, using the equation, purely in terms of initial data. First, note that equation \eqref{FutureRescWaveEq} can be written in terms of variables $(t,r,\omega )$ as
\[ (\partial_t + \partial_r ) (\partial_t - \partial_r ) \hat\phi  - \frac{1}{r^2} \Delta_{S^2} \hat\phi = 0 \, .\]
Whence,
\begin{eqnarray*}
\partial_R \left( \begin{array}{c} \hat\phi \\ \partial_t \hat\phi \end{array} \right) &=& -r^2 \left( \partial_t + \partial_{r} \right) \left( \begin{array}{c} \hat\phi \\ \partial_t \hat\phi \end{array} \right) \\
&=& -r^2 \left( \begin{array}{cc} {\partial_{r}} & 1 \\ {\partial_{r}^2 + \frac{1}{r^2} \Delta_{S^2}} & {\partial_{r}} \end{array} \right) \left( \begin{array}{c} \hat\phi \\ \partial_t \hat\phi \end{array} \right) =: L \left( \begin{array}{c} \hat\phi \\ \partial_t \hat\phi \end{array} \right) \, .
\end{eqnarray*}
The operator $L$ purely involves spacelike derivatives. We can now express the spaces of initial data that entail peeling at a given order.
\begin{definition} \label{PeelingSpaces}
Given $\hat\phi_0,\hat\phi_1 \in {\cal C}^\infty_0 (\left[ -u_0 ,+\infty \right[_{r_*} \times S^2_\omega )$, we define the following squared norm of order $k$~:
\begin{equation}
\left\| \left( \begin{array}{c} \hat\phi_0 \\ \hat\phi_1 \end{array} \right) \right\|_k^2 := \sum_{p=0}^k \sum_{q=0}^p {\cal E}_{T,{\Sigma}_{0}^{u_0}} \left( L^q \nabla_{S^2}^{p-q} \left( \begin{array}{c} \hat\phi_0 \\ \hat\phi_1 \end{array} \right) \right) \, , \label{OrderkNorm}
\end{equation}
where we have denoted by ${\cal E}_{T,{\Sigma}_{0}^{u_0}} \left( \begin{array}{c} \hat\phi_0 \\ \hat\phi_1 \end{array} \right)$ the energy ${\cal E}_{T,{\Sigma}_{0}^{u_0}} (\hat\phi )$, given in \eqref{EnergySigma0}, where $\hat\phi$ is replaced by $\hat\phi_0$ and $\partial_t \hat\phi = \partial_u \hat\phi$ is replaced by $\hat\phi_1$.
\end{definition}
\begin{theorem}
The space of initial data (on $\left[ -u_0 ,+\infty \right[_{r} \times S^2_\omega$) for which the associated solution peels at order $k$ is the completion of ${\cal C}^\infty_0 (\left[ -u_0 ,+\infty \right[_{r} \times S^2_\omega ) \times {\cal C}^\infty_0 (\left[ -u_0 ,+\infty \right[_{r} \times S^2_\omega )$ in the norm (\ref{OrderkNorm}). The fact that we have estimates both ways at all orders guarantees that this setting is optimal for our definition.
\end{theorem}

\subsection{Results on the Schwarzschild metric} \label{SchwaPeel}

The Schwarzschild metric expressed in terms of Schwarzschild coordinates is
$$
g=\left( 1-\frac{2m}r \right) \d t^2- \left( 1-\frac{2m}r\right)^{-1} \d r^2 -r^2\d\omega^2
$$
where $m>0$ is the mass of the black hole and $\d\omega^2$ is the Euclidean metric on the unit sphere $S^2$. We work on the exterior of the black hole
\[ {\cal M} = \R_t \times ]2m , +\infty [_r \times S^2_\omega \, .\]
The associated d'Alembertian is
\[ \square_g = \left( 1 -\frac{2m}{r} \right)^{-1} \frac{\partial^2}{\partial t^2} - \frac{1}{r^2} \frac{\partial}{\partial r} \, r^2 \left( 1 -\frac{2m}{r} \right) \frac{\partial}{\partial r} - \frac{1}{r^2} \Delta_{S^2} \, ,\]
where $\Delta_{S^2}$ is the Laplacian on $S^2$ endowed with the Euclidean metric. We perform a conformal rescaling that is similar to the partial compactification of Minkowski spacetime~: we consider the variables
$$
R=1/r \, ,~ u=t-r_* \, ,~\mathrm{with~} r_*=r+2m\log(r-2m)\, ,
$$
and rescale the metric $g$ as follows
$$
\hat{g}= R^2 g= R^2(1-2mR)\d u^2-2\d u\d R-\d\omega^2 \, .
$$
In these coordinates,
\[ \scri^+ = \{ 0 \}_R \times \R_u \times S^2_\omega \, .\]
The scalar curvature of $\hat{g}$ is given by
\[ \mathrm{Scal}_{\hat{g}} =12mR \]
and the conformally invariant wave equation on the metric $\hat{g}$ has the following form
\begin{equation} \label{SchwaConfWaveEq}
\left( \square_{\hat{g}} + 2mR \right) \hat\phi = \left( -2 \partial_u \partial_R - \partial_R R^2 \left( 1-2mR
\right) \partial_R - \Delta_{S^2} +2mR \right) \hat\phi = 0 \, .
\end{equation}
As in the case of flat spacetime, the two following properties are equivalent~:
\begin{enumerate}
\item $\phi \in {\cal D}' ({\cal M})$ satisfies $\square_g \phi =0$\, ;
\item $\hat\phi := R^{-1} {\phi} = r{\phi}$ satisfies (\ref{SchwaConfWaveEq}) on ${\cal M}$.
\end{enumerate}
The situation is now slightly more complicated because equation \eqref{SchwaConfWaveEq} does not admit a conserved stress-energy tensor. We choose to use the stress-energy tensor for the free wave equation
\begin{equation} \label{SETCWESchwa}
\hat{T}_{ab}=\hat{T}_{(ab)}=\partial_a \hat\phi \, \partial_b \hat\phi - \frac12 \hat{g}_{ab} \hat{g}^{cd} \partial_c \hat\phi \, \partial_d \hat\phi \, ,
\end{equation}
which satisfies
\begin{equation} \label{ConsLawSETCWESchwa}
\nabla^a \hat{T}_{ab} = \square_{\hat{g}} \hat\phi \, \partial_b \hat\phi = - 2mR \hat\phi \, \partial_b \hat\phi\, .
\end{equation}
Then we adapt the flat spacetime Morawetz vector field to construct a timelike vector field transverse to $\scri^+$~: this is a classic construction in the litterature (it was first introduced by Inglese and Nicolò \cite{InglNi} and then used by other authors, for instance by Dafermos and Rodnianski in \cite{DaRo}), one chooses a coordinate system in Minkowski spacetime, obtains the expression of the Morawetz vector field in these coordinates, then keeps the expression in an analogous coordinate system on the Schwarzschild spacetime. So in a sense, there are several Morawetz vector fields on $\cal M$, depending on the coordinate system one chooses. The usual choice is to work with $u=t-r$, $v=t+r$, $\omega$ and then to transpose the expression in the coordinates $u=t-r_*$, $v=t+r_*$, $\omega$ on $\cal M$. Instead, we use the expression \eqref{MorawetzVF} and define on $\cal M$ the vector field
\begin{equation} \label{SMVF} 
T := u^2 \partial_u -2 (1+uR) \partial_R
\end{equation}
in the coordinate system $u = t-r_*$, $R=1/r$, $\omega$. This is now no longer a Killing vector field, but its Killing form has a rather fast decay at infinity~:
\[ \hat\nabla_{(a} T_{b)} \d x^a \d x^b = 4mR^2(3+uR)\d u^2 \, .\]
The associated energy current will satisfy an approximate conservation law, with error terms coming both from the equation and the Killing form of $T$. More precisely,
\[ J^a := T_b \hat{T}^{ab} \, ;~ \hat\nabla_{a} J^a = - 2mR \nabla_T \hat\phi + \nabla_{(a} T_{b)}\hat{T}^{ab} \, . \]
Then, it simply remains to apply the method we have developed in the flat case and to check that all error terms can be controlled via a priori estimates of Gronwall type. The details can be found in \cite{MaNi2009} for the wave equation and \cite{MaNi2012} for Dirac and Maxwell fields. We now express the definition of the peeling at any order on the Schwarzschild metric and its characterization in terms of classes of initial data. As before, we work on a domain $\{ u \leq u0 \}$ for $u_0 << -1$ and we consider the energy fluxes through the three parts of its boundary (with the same notations as in the flat case)
\begin{eqnarray}
{\cal E}_{T,{\Sigma}_{0}^{u_0}} (\hat\phi ) &=& \int_{{\Sigma}_{0}^{u_0}} \left( u^2 (\partial_u \hat\phi )^2 + R^2(1-2mR) u^2 \partial_u \hat\phi \partial_R \hat\phi \right. \nonumber \\
&& + R^2(1-2mR) \left( \frac{(2+uR)^2}{2} -mu^2 R^3 - (1+uR)
\right) (\partial_R \hat\phi)^2 \nonumber \\
&&  \left. + \left( \frac{u^2 R^2 (1-2mR)}{2} + 1+uR \right) \left| \nabla_{S^2} \hat\phi \right|^2 \right) \d u \d^2 \omega \nonumber \\
& \simeq & \int_{{\Sigma}_{0}^{u_0}} \left( u^2(\partial_u \hat\phi )^2 + \frac{R}{|u|} (\partial_R \hat\phi )^2+ \left| \nabla_{S^2} \hat\phi \right|^2 \right)  \d u\d^2\omega \, , \nonumber \\
{\cal E}_{T, \scri^+_{u_0}} (\hat{\phi}) &=& \int_{\scri^+_{u_0}} \left( u^2(\partial_u \hat\phi )^2 + \vert\nabla_{S^2} \hat\phi \vert^2 \right) \d u \d^2 \omega \, , \nonumber \\
{\cal E}_{T,{\cal S}_u} (\phi ) &=& \int_{{\cal S}_u} \frac12\left( \left( (2+uR)^2 - 2mu^2R^3 \right)(\partial_R \hat\phi )^2+u^2 \vert\nabla_{S^2}\hat\phi \vert^2 \right) \d R \d ^2 \omega \, . \nonumber
\end{eqnarray}
We obtain estimates \eqref{MinkHOEst1} and \eqref{MinkHOEst2} in the Schwarzschild framework and the spherical symmetry once again allows us to get analogous estimates for angular derivatives. Hence the following definition of the peeling at any order on the Schwarzschild metric~:
\begin{definition}
We say that a solution $\hat\phi$ of (\ref{SchwaConfWaveEq}) peels at order $k\in \N$ if for all polynomials $P$ in $\partial_R$ and $\nabla_{S^2}$ of order lower than or equal to $k$, we have ${\cal E}_{T,\scri^+_{u_0} } (P \hat\phi ) <+\infty$. This means than for all $p \in \{ 0,1,...,k\}$ we have for all $q \in \{ 0,1,...,p \}$, ${\cal E}_{T,\scri^+_{u_0} } (\partial_R^q \nabla_{S^2}^{p-q} \hat\phi ) <+\infty$. This condition can be re-expressed as the finiteness of the following norm of the data for $\hat\phi$ at $t=0$~:
\begin{equation}
\left\Vert \left( \begin{array}{c} {\hat\phi_0} \\ {\hat\phi_1} \end{array} \right) \right\Vert_k^2 := \sum_{p=0}^k \sum_{q=0}^p {\cal E}_{T,{\Sigma}_{0}^{u_0}} \left( L^q \nabla_{S^2}^{p-q} \left( \begin{array}{c} \hat\phi_0 \\ \hat\phi_1 \end{array} \right) \right) \, , \label{SchwaOrderkNorm}
\end{equation}
the operator $L$ now reading
\[ L = -\frac{r^3}{r-2m} \left( \begin{array}{cc} {\partial_{r_*}} & 1 \\ {\partial_{r_*}^2 - \frac{2m(r-2m)}{r^4} + \frac{r-2m}{r^3} \Delta_{S^2}} & {\partial_{r_*}} \end{array} \right)  \, .\]
\end{definition}
\begin{theorem}
The space of initial data (on $\left[ -u_0 ,+\infty \right[_{r_*} \times S^2_\omega$) for which the associated solution peels at order $k$ is the completion of ${\cal C}^\infty_0 (\left[ -u_0 ,+\infty \right[_{r_*} \times S^2_\omega ) \times {\cal C}^\infty_0 (\left[ -u_0 ,+\infty \right[_{r_*} \times S^2_\omega )$ in the norm (\ref{SchwaOrderkNorm}). This is optimal for our definition.
\end{theorem}
The remaining task is to compare the characterizations of peeling at the same given order between the flat case and the Schwarzschild case. After the care we took to make sure that our constructions would be as close to one another as possible, it turns out that not only is comparison between our classes of data natural, but also these classes are almost trivially equivalent in the following sense~: the classes are defined by weighted Sobolev norms~; the weights intervening in the flat and Schwarzschild cases have equivalent behaviours at infinity when using $r$ for the radial variable in $\mathbb{M}$ and $r_*$ in $\cal M$\footnote{In fact, the choice of $r_*$ in the Schwarzschild situation is not crucial for the comparison of the asymptotic behaviour of the weights, simply because $r_* \simeq r$ at infinity. This choice is however the natural one to make because the radial derivatives appearing in the norms are $\partial_r$ on $\mathbb{M}$ and $\partial_{r_*}$ on $\cal M$.}. The essential reason for this is that the norms in the Schwarzschild situation are uniform in the mass $m$ of the black hole on any given bounded interval $]0,M]$. The details can be found in \cite{MaNi2009,MaNi2012}.

\section{Conformal scattering} \label{CS}

Scattering theory is a way of summarizing the whole evolution of solutions to a certain equation by a scattering operator that, to their asymptotic behaviour in the distant past, associates their asymptotic behaviour in the distant future. These asymptotic behaviours are usually solutions to a simpler equation, a comparison dynamics. A complete scattering theory will not only show the existence of a scattering operator but will also establish that the solutions are completely and uniquely characterized by their past (resp. future) asymptotic behaviours, which entails in particular the invertibility of the scattering operator. Different choices of comparison dynamics are always possible, giving different scattering operators that are not necessarily defined on the same function spaces. Some simplified dynamics can be described geometrically as transport equations along congruences of null geodesics defining null infinity~; the fields that they propagate then correspond to functions defined on $\scri$. Using such comparison dynamics for scattering theories on asymptotically flat spacetimes means that the scattering data, i.e. the large time asymptotic behaviours, are merely radiation fields.

The idea of using a conformal compactification in order to obtain a time-dependent scattering theory formulated in terms of radiation fields is due to Roger Penrose. His discussion of the topic in \cite{Pe1965} clearly indicates that it was one of the main motivations of the conformal technique. The first actual conformal scattering theory appeared fifteen years later in F.G. Friedlander's founding paper \cite{Fri1980} as a combination of F.G. Friedlander's own work on radiation fields \cite{Fri1962,Fri1964,Fri1967} and the Lax-Phillips approach to time-dependent scattering \cite{LaPhi}. This first paper was treating the case of the conformal wave equation. The geometrical background was a static asymptotically flat spacetime with a fast decay at infinity, too fast for allowing for the presence of energy when considering solutions of the Einstein vacuum equations, but fast enough to ensure a smooth conformal compactification including at spacelike and timelike infinities. The principle of the construction was first to re-interpret the scattering theory as the well-posedness of the Goursat problem for the rescaled equation at null infinity, then to solve this Goursat problem.  F.G. Friedlander's main goal was then, it seems to me, to extend the results of Lax and Phillips, in all their analytic precision, to a curved situation. In particular, he wanted to recover the fundamental structure in the Lax-Phillips theory~: a translation representer of the solution. The existence of such an object requires a timelike Killing vector field that extends as the null generator of $\scri$. This is probably what motivated the choice of a static background, even though the conformal scattering construction itself can be performed on non stationary geometries. F.G. Friedlander's method was then applied to nonlinear equations, but still for static geometries, by J.C. Baez, I.E. Segal and Zhou Z.F. in 1989-1990 \cite{Ba1989a, Ba1989b, Ba1990, BaSeZho1990, BaZho1989}. F.G. Friedlander himself at the end of his life came back to conformal scattering in a posthumously published note \cite{Fri2001}. L. Hörmander published in 1990 a short paper, in the form of a remark prompted by \cite{BaSeZho1990}, entitled ``A remark on the characteristic Cauchy problem'' \cite{Ho1990}, in which he presented a general method for solving the initial value problem on a weakly spacelike hypersurface, for a general wave equation on spatially compact spacetimes. The method is entirely based on energy estimates and compactness methods. Using this approach, a conformal scattering theory on generically non stationnary backgrounds was developed by L.J. Mason and the author \cite{MaNi2004}. Then, more recently, J. Joudioux obtained the first result for a non linear wave equation \cite{Jo2012} in non stationary situations. I came back to the topic a couple of years ago to propose an extension of these methods to black hole spacetimes \cite{Ni2013}, describing the construction in the Schwarzschild spacetime, which is static, and discussing the case of the Kerr metric and the associated difficulties.

This section starts by a brief description of the Lax-Phillips theory in a simplified setting. Then we describe the conformal scattering construction for the wave equation in the flat case and move on to extensions to non stationary and black hole situations.

\subsection{A simple overview of Lax-Phillips theory}

The Lax-Phillips theory describes the scattering of a massless scalar field by an obstacle. We present here a version of the theory without obstacle, i.e. for the free wave equation. Spectral analysis is used to construct a translation representer of the free wave equation, which is then re-interpreted geometrically as a radiation field. We describe the construction of the translation representer and its geometrical reinterpretation. It is usual to consider that in the case of a free equation, there is no scattering. Indeed the existence of the translation representer can be understood in this manner. But this statement is not invariant, it merely corresponds to chosing the equation itself as comparison dynamics. The geometrical re-interpretation of the translation representer describes asymptotic behaviours as radiation fields. For this choice of comparison dynamics, the scattering process is non trivial even on flat spacetime. Moreover the resulting scattering operator can be defined geometrically without resorting to the spectral analytic part of the Lax-Phillips theory. The construction thus modified can be easily generalized to a large class of curved situations. This will be the object of subsections \ref{CScatMink}, \ref{CScatASST} and \ref{CScatSchwa}.

\subsubsection{Finite dimensional case}

Let us first describe the essential ideas on a finite dimensional toy model. Consider the following equation for a time-dependent vector in $\C^n$~:
\begin{equation} \label{DiffSys}
\partial_t V (t) = iA V(t)
\end{equation}
where $A$ is an $n\times n$ hermitian matrix $A$ with $n$ distinct eigenvalues $\sigma_1$, ..., $\sigma_n$. Let $\{ e_1 \, ,~ ... \, ,~ e_n \}$ be an orthonormal basis of eigenvectors of $A$. The Cauchy problem for \eqref{DiffSys} is solved by the propagator $e^{itA}$, i.e. if $V$ is a solution of \eqref{DiffSys},
\[ V(t) = e^{i(t-s)A} V(s) \, ,~ \forall t,s\in \R \]
and the matrices $e^{itA}$ are unitary.

Instead of considering $V$ as an element of $\C^n$, we represent it as a function on the spectrum of $A$, which is square integrable for the natural spectral measure $\mu = \sum_{i=1}^n \delta_{\sigma_i}$~:
\[ \begin{array}{ccccc} {\C^n} & \rightarrow & {L^2 (\R \, ;~ \d \mu)}  \\
V & \mapsto & {\tilde{V} (\sigma_i ) = \langle V ,  e_i \rangle \, .} \end{array} \]
This is a spectral representation in the sense that the action of $A$ is now described simply by multiplication by the spectral parameter~:
\[ \widetilde{A V} (\sigma) = \sigma \tilde{V} (\sigma) \, .\]
Then taking the Fourier transform,
\[ \hat{\tilde{V}} := {\cal F}_\sigma \tilde{V} \, , \]
we obtain a new representation for which the evolution is described by a simple translation of $t$~:
\[ {\cal F}_\sigma \left( \widetilde{e^{itA} V} \right) (s) = {\cal F}_\sigma (e^{it\sigma} \tilde{V} ) (s) = \hat{\tilde{V}} (s-t) \, . \]
This is called a translation representer of the solution of \eqref{DiffSys}. A similar construction can be performed for the wave equation on Minkowski spacetime and is at the heart of the Lax-Phillips theory.

\subsubsection{The wave equation}

We now consider the wave equation on Minkowski spacetime~:
\begin{equation} \label{WE}
\partial_t^2 \phi - \Delta \phi =0 \, .
\end{equation}
It can be written formally as equation \eqref{DiffSys} in the following manner~:
\[ \partial_t U = iA U \, ,~ U := \left( \begin{array}{c} \phi \\ \partial_t \phi \end{array} \right) \, ,~ A = -i\left( \begin{array}{cc} 0 & 1 \\ \Delta & 0 \end{array} \right) \, .\]
The operator $A$ is self-adjoint on ${\cal H} = \dot{H}^1 (\R^3 ) \times L^2 (\R^3 )$, where $\dot{H}^1 (\R^3)$ is the first order homogeneous Sobolev space on $\R^3$, completion of ${\cal C}^\infty_0 (\R^3)$ in the norm
\[ \Vert \psi \Vert^2_{\dot{H}^1 (\R^3)} = \int_{\R^3} \vert \nabla \psi \vert^2 \d^3 x \, . \]
The spectrum of $A$ is the whole real axis and is purely absolutely continuous. In particular the point spectrum of $A$ is empty. The equation $AU=\sigma U$ for $\sigma \in \R$, which reads
\begin{equation}
\left\{ \begin{array}{rcl} { u_2} & = & {i \sigma u_1 \, ,} \\ {\Delta u_1} &=& {i \sigma u_2 \, ,} \\
&=& {-\sigma^2 u_1 \, ,} \end{array} \right.
\end{equation}
does not have any finite energy solution, by which we mean a solution in $\cal H$. However, for each $\sigma \in \R^*$, $A$ has a whole $2$-sphere worth of generalized eigenfunctions which are plane waves~:
\[ e_{\sigma , \omega} (x) = \left( \begin{array}{c} {e^{-i \sigma x.\omega}} \\ { i \sigma e^{-i \sigma x.\omega}} \end{array} \right) \, ,~ \omega \in S^2 \, .\]
For $\sigma = 0$, the $2$-sphere collapses to a point and the only solution is
\[ e_{0,\omega} (x) = e_{0} (x) = \left( \begin{array}{c} {1} \\ {0} \end{array} \right) \, . \]
We now proceed exactly as in the finite dimensional case. Consider $U \in {\cal C}^\infty_0 (\R^3 ) \times {\cal C}^\infty_0 (\R^3 )$, we represent it as a function on $\R_\sigma \times S^2_\omega$ by taking its inner product with plane waves, suitably normalized~:
\begin{eqnarray*}
\tilde{U} (\sigma , \omega ) &:=& \frac{1}{(2\pi )^{3/2}} \langle U , e_{\sigma , \omega} \rangle_{\cal H} \\
&=&  \frac{1}{(2\pi )^{3/2}} \int_{\R^3} ( \nabla u_1 \overline{\nabla e^{-i \sigma x.\omega}} + u_2 \overline{i \sigma e^{-i \sigma x.\omega}} ) \d^3 x \\
&=& \frac{1}{(2\pi )^{3/2}} \int_{\R^3} ( u_1 (\overline{-\Delta e^{-i \sigma x.\omega}}) + u_2 \overline{i \sigma e^{-i \sigma x.\omega}} ) \d^3 x \\
&=&  \frac{1}{(2\pi )^{3/2}} \int_{\R^3} ( \sigma^2 u_1 - i \sigma u_2 ) e^{i \sigma x.\omega} \d^3 x \, \\
&=& \sigma^2 \hat{u}_1 (-\sigma \omega ) -i \sigma \hat{u}_2 (-\sigma \omega ) \, ,
\end{eqnarray*}
where ``$\, \hat{~}\, $'' denotes the Fourier transform on $\R^3$. Although the intermediate calculations do not, the final formula extends to $\cal H$ and the map that to $U$ associates $\tilde{U}$ extends as an isometry from $\cal H$ onto $L^2 (\R_\sigma \times S^2 )$. This provides a spectral representation of $A$ and its propagator~:
\[ \widetilde{AU} = \sigma \tilde{U} \, ,~ \widetilde{e^{itA} U} = e^{it\sigma} \tilde{U} \, .\]
Then we take the Fourier transform in $\sigma$ to obtain the new representation $\cal R$~:
\[ {\cal R} U (r,\omega ) := {\cal F}_\sigma (\tilde{U} (.,\omega)) ( r) \, .\]
Then just as in the finite dimensional case, $\cal R$ is a translation representation~; we have
\[ {\cal R} ( e^{itA} U ) (r, \omega) = ({\cal R} U ) (r-t , \omega ) \, .\]
This representation is of course also an isometry from $\cal H$ onto $L^2 (\R \times S^2 )$.

In a sense, the existence of a translation representer can be interpreted as the fact that there is no scattering~; the evolution merely corresponds to the translation, without deformation, of a function representing the initial data. The Lax-Phillips theory does not stop there however. The representation $\cal R$ can be expressed as follows
\[ {\cal R} U = \frac{1}{4\pi } (-\partial_s^2 Ru_1 + \partial_s Ru_2 ) (s,\omega )  \, ,\]
where $R$ is the Radon transform $R$ defined for $f \in {\cal C}^\infty_0 (\R^3 )$ by
\[ Rf (s,\omega ) = \int_{ x.\omega = s } f(x) \d^2 \sigma( x) \, .\]
This observation and the knowledge of the inverse Radon transform
\[ R^* \psi (x) = \int_{S^2} \psi (x.\omega , \omega ) \d^2 \omega \]
give an explicit converse map to $\cal R$~:
\[ {\cal I} k = \frac{1}{2\pi} (R^* k \, ,~ - R^* \partial_s k ) \, .\]
Lax and Phillips then use this to establish the {\it asymptotic profile} property which essentially re-interprets the translation representer as a future radiation field~:
\begin{equation} \label{AsP}
{\cal R} U (s,\omega ) = - \lim_{r\rightarrow +\infty} r \partial_t \phi (r , (r+s)\omega ) \, .
\end{equation}
In addition, the fact that $\cal I$ is the inverse of $\cal R$ gives immediately an integral formula for the solution in terms of its translation representer
\[ \phi (t,x) = \frac{1}{2\pi} \int_{S^2} {\cal R}U (x.\omega +t , \omega ) \d^2 \omega \, . \]
This is precisely Whittaker's formula from 1903 \cite{Whi1903} for the solutions to the wave equation, which the Lax-Phillips theory allowed to reinterpret as providing the solution to the Goursat problem for the wave equation at null infinity.

Performing a similar construction for $-A$ instead of $A$, we obtain new representations $\check{\cal R}$ and $\check{\cal I}$ which relate the solution to its past radiation field. The scattering operator, which to the past radiation field associates the future radiation field, is then given by
\[ S = {\cal R} \check{\cal I} \, .\]

\subsection{Conformal scattering on Minkowski spacetime} \label{CScatMink}

The Lax Phillips theory provides a scattering operator turning past radiation fields into future radiation fields and which is therefore naturally understood as acting on compactified Minkowski spacetime. However the construction leading to this operator is entirely performed on $\mathbb{M}$, not on the compactified spacetime. Besides the techniques used require a static, or at least a stationary, background. We provide here an alternative construction of a similar scattering operator described in terms of the radiation field for $\phi$ instead of that for $\partial_t \phi$. Our construction is performed on the compactified spacetime using essentially the regularity of the conformal boundary. It follows the approach of \cite{Ho1990} for the resolution of the Goursat problem and is done in three main steps. We adopt the full compactification of Minkowski spacetime and the notations of section \ref{FullCompact}, i.e. the rescaled field is denoted $\tilde\phi$.

\begin{description}
\item[Step 1.] We define the trace operator $T^+$ that to the data for $\tilde\phi$ at $\tau=0$ associates the trace of $\tilde{\phi }$ on $\scri^+$. This operator is well defined for data that extend as smooth functions on $S^3$, in particular for $\tilde{\phi}_0$, $\tilde{\phi}_1 \in {\cal C}^\infty_0 (\R^3)$. The image of such data is a smooth function on $\scri^+$\footnote{Note that in the case of Minkowski spacetime, we could straight away consider data $\tilde\phi_0$, $\tilde\phi_1 \in {\cal C}^\infty (S^3)$ to define $T^+$. This cannot be extended to more general situations however. Hence we prefer to use a more flexible approach that is very common in scattering theory~: to start with smooth compactly supported data and extend the operators by density using uniform estimates.}.
\item[Step 2.] We prove energy estimates both ways between the data and their image through $T^+$. In the case of Minkowski spacetime, we have a stronger result which is the energy equality \eqref{EnEqEin}, saying that for $\tilde\phi_0$, $\tilde\phi_1  \in {\cal C}^\infty (S^3) $,
\[ {\cal E}_{K,\scri^+} (\tilde{\phi} ) = {\cal E}_{K,X_0} (\tilde{\phi} ) \, . \]
This implies that $T^+$ extends as a partial isometry from $H^1 (S^3 ) \times L^2 (S^3)$ into $H^1 (\scri^+ )$ (see \eqref{NormS3} and \eqref{NormScri} for the energy norms on $X_\tau$ and $\scri^+$) and is one-to-one with closed range.
\item[Step 3.] In order to prove that the trace operator is onto, we only need to establish that its range is dense in $H^1 (\scri^+ )$ since we already know that it is a closed subspace of $H^1 (\scri^+ )$. This is done by solving the Goursat problem for a dense subset of $H^1 (\scri^+ )$. We must be able to find a solution of the rescaled equation for which we have access to the trace on $\scri^+$ in the strong sense and to check that this trace is the data we started from. For the Goursat problem on $\scri^+$ for the rescaled wave equation on the Einstein cylinder, this follows from \cite{Ho1990} for data in $H^1 (\scri^+)$, so we prove directly the surjectivity without needing to use the closed range property.
\end{description}
After these three steps, we have established that the operator $T^+$ is an isometry from $H^1 (S^3) \times L^2 (S^3)$ onto $H^1 (\scri^+)$. A similar construction can be performed in the past for the trace operator $T^-$. Then the scattering operator that maps past radiation fields to future radiation fields\footnote{Here we call radiation fields the traces of the rescaled solution at $\scri^\pm$. This is not quite the way the radiation fields were defined in \eqref{FRF} and \eqref{PRF}. The two notions of radiation fields only differ by the multiplication by fixed functions on $\scri^\pm$, which are the limits at $\scri^\pm$ of the ratio of the two conformal factors $1/r$ and $\Omega$. We shall use in this section the more flexible definition as traces of the rescaled field for the conformal factor we choose to work with.} is simply given by
\[ S := T^+ (T^-)^{-1} \, .\]
The above construction does not use any of the symmetries of Minkowski spacetime. All it requires is a regular conformal compactification. This is in fact quite strong. When dealing with asymptotically flat solutions to the Einstein vacuum equations, one does not expect that the conformal metric will be smooth at $i^0$ unless the ADM mass is zero and the spacetime is flat. In the case of black hole spacetimes, the situation is even worse since timelike infinities are rather strong singularities of the conformal metric. The method needs to be modified in order to deal with these singularities. The treatment of the difficulty at $i^0$ can be found in \cite{MaNi2004} and a conformal scattering on the Schwarzschild metric was developed in \cite{Ni2013} with a way of dealing with timelike infinities. We give in the next two subsections the essential ingredients of the extension on the conformal scattering construction to spacetimes with singular $i^0$ and to black hole spacetimes.

\subsection{The case of asymptotically simple spacetimes} \label{CScatASST}

\subsubsection{Geometrical background}

We work on smooth globally hyperbolic asymptotically simple spacetimes that contain energy (the ADM energy is not zero) and such that $i^\pm$ are regular points of the conformal structure. The fact that the ADM energy does not vanish means in particular that the conformal structure is singular at $i^0$. More precisely, the type of spacetimes we work on are as follows~: a
4-dimensional, globally hyperbolic, Lorentzian space-time $\left(
  \scrm, g\right)$, $\scrm \simeq \R^4$, such that there exists another globally hyperbolic,
Lorentzian space-time $(\scrmhat , \hat{g} )$ and a smooth scalar
function $\Omega$ on $\scrmhat$ satisfying~:
\begin{description}
\item[(i)] $\scrm$ is the interior of $\scrmhat$, its boundary is the union of two points $i^-$ and $i^+$ and a smooth
  null hypersurface $\scri$, which is the disjoint union of the past light-cone
  $\scri^+$ of $i^+$ and of the future light-cone $\scri^-$ of $i^-$~;
\item[(ii)] $\Omega >0$, $\hat{g}= \Omega^2 g$ on $\scrm$,
  $\Omega =0$ and $\d \Omega \neq 0$ on $\partial \scrm$~;
\item[(iii)] every inextendible null geodesic in $\scrm$ acquires a future
  endpoint on $\scri^+$ and a past endpoint on $\scri^-$.
\end{description}
Since null unparametrized geodesics are conformally invariant objects and since from any point of $\scri$ we can find a null geodesic for $\hat{g}$ that enters the spacetime, the definition above implies that $\scri^+$ (resp. $\scri^-$)
is the set of future (resp. past) end-points of null geodesics. Spacelike infinity, $i^0$, is the boundary of Cauchy hypersurfaces in $\scrm$~; it does not belong to $\scrmhat$ therefore the definition remains somewhat abstract but we shall not need to make it more precise here.

In the results that have appeared sofar (i.e. \cite{Jo2012} and \cite{MaNi2004}), the following additional symmetry assumption was imposed~:

\noindent
{\bf (iv)} $\scrm$ is diffeomorphic to Schwarzschild's spacetime outside the domain of influence of a given compact subset $K$ of a Cauchy hypersurface $\Sigma_0$. In particular, we assume that outside the domain of influence of $K$, $\Omega = 1/r$ where $r$ is the radial variable in the Schwarzschild coordinates.

The reason for this additional hypothesis is that {\bf (i)}-{\bf(iii)} do not give enough control on the geometry of $(\scrmhat , \hat{g} )$ near $i^0$ to perform energy estimates in that region. In order to gain this control, we can either impose some symmetry, or specify the decay in spacelike and null directions of $\hat{g}$ and its derivatives of order up to $2$. The first solution has the advantage of simplicity. Besides, there are large classes of solutions to the Einstein vacuum equations whose behaviour is generically non stationary within the domain of influence of $K$ and which satisfy assumptions {\bf (i)}-{\bf(iv)} (see \cite{ChruDe2002,ChruDe2003,Co2000,CoScho2003}).

\subsubsection{Dealing with spacelike infinity}

On the class of asymptotically simple spacetimes defined above, we adopt the same strategy as on Minkowski spacetime in order to construct a conformal scattering theory. We describe the construction towards the future. A similar one can be performed towards the past and, put together, they provide the scattering operator. We choose a Cauchy hypersurface $\Sigma_0$ and denote by $\hat{\scrm}^+$ the part of $\hat{\scrm}$ in the future of $\Sigma_0$.
\begin{description}
\item[Step 1.] We take smooth compactly supported initial data on $\Sigma_0$~: $\hat{\phi}_0$, $\hat{\phi}_1 \in {\cal C}^\infty_0 (\Sigma_0 )$ and consider $\hat{\phi} \in {\cal C}^\infty (\scrm )$ the associated solution of
\begin{equation}\label{RescEq}
\left( \square_{\hat{g}} + \frac16 \mathrm{Scal}_{\hat{g}} \right) \hat{\phi} =0 \, .
\end{equation}
As a simple consequence of the finite propagation speed (which implies that the support of $\hat{\phi}$ remains away from $i^0$) and of the regularity of $\hat{g}$ up to the boundary of $\scrmhat$, $\hat\phi$ extends as a smooth function on $\scrmhat$. We can therefore define the trace operator $T^+$ that to data $\hat{\phi}_0$, $\hat{\phi}_1 \in {\cal C}^\infty_0 (\Sigma_0 )$ associates the trace of the associated solution $\hat\phi$ on $\scri^+$.
\item[Step 2.] We establish estimates both ways between the energy of the data and that of their image through $T^+$. First, we need to choose a timelike vector field with which to define the energies. We use the symmetry assumption {\bf (iv)}. Outside the domain of influence of $K$, we have a Morawetz vector field associated to the Schwarzschild metric by the construction done in subsection \ref{MinimalCompact} and we extend it as a smooth timelike vector field $T^a$ over $\scrmhat$ (the decomposition of the future of $\Sigma_0$ into the domain of influence of $K$ and its complement as well as the construction of $T^a$ are shown in figure \ref{PASST}).
\begin{figure}[htb!]
\centering
  \begin{tabular}{@{}cc@{}}
   \hspace{-0.1in}
   \includegraphics[width=.5\textwidth]{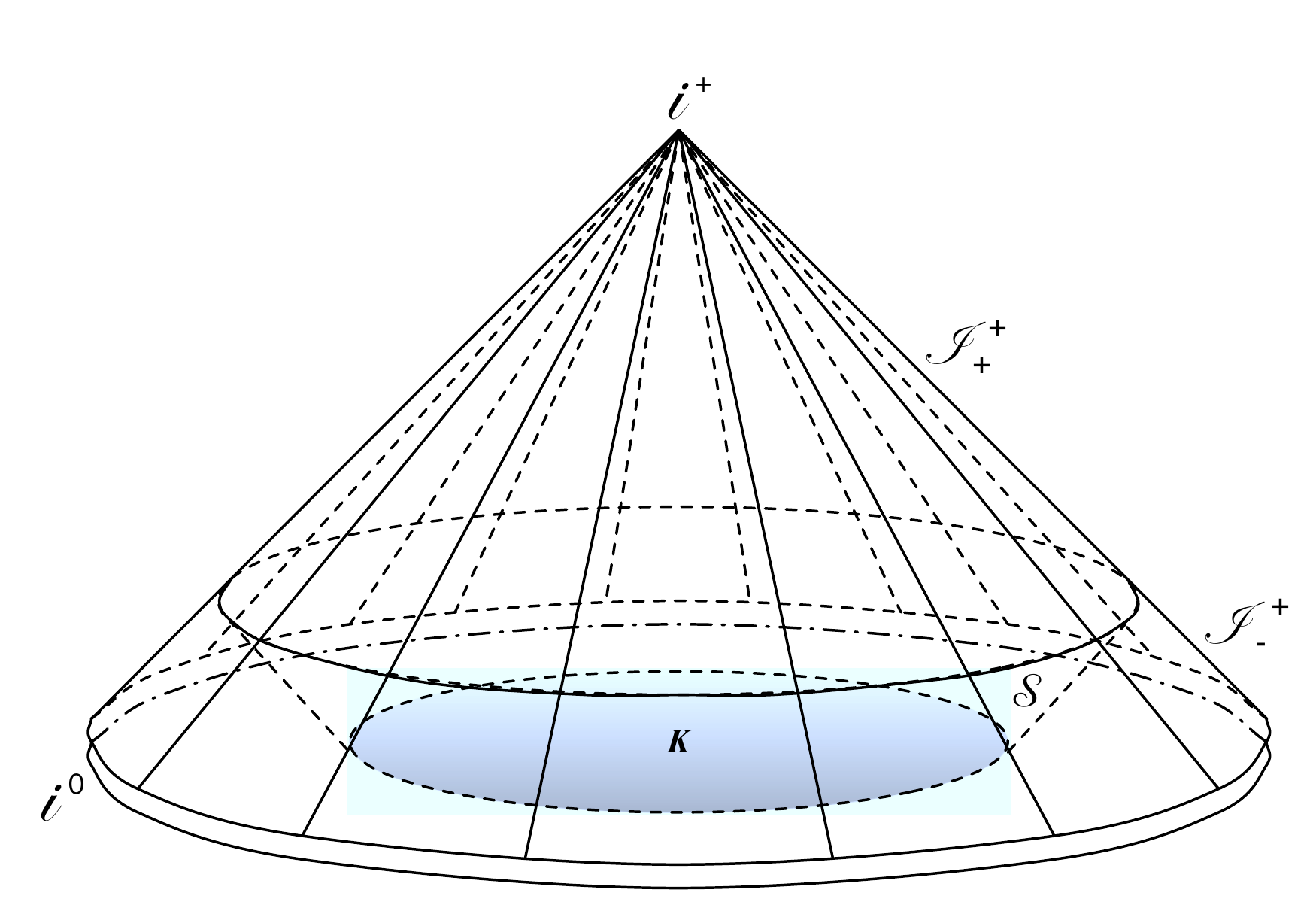} &
   \hspace{-0.1in}
   \includegraphics[width=.5\textwidth]{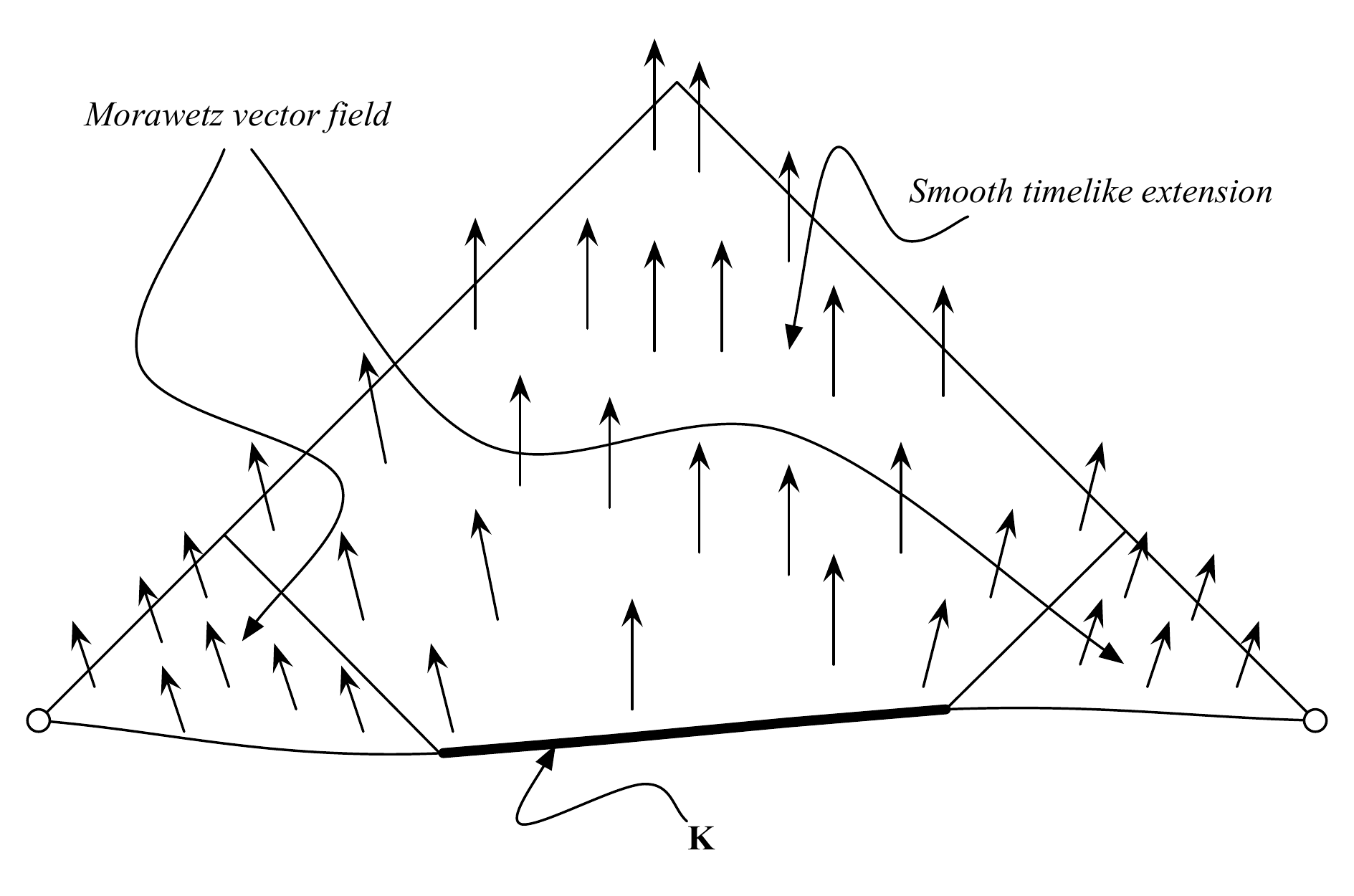}
  \end{tabular}
  \caption{Future of $\Sigma_0$ in $\hat{\scrm}$, then a 2D-cut with a choice of global timelike vector field.} \label{PASST}
\end{figure}
Let us denote by $\scrs$ the boundary of the future domain of influence of $K$ in $\scrm$, by $\scri^+_+$ the part of $\scri^+$ in the future of $\scrs$ and by $\scri^+_-$ the part of $\scri^+$ in the past of $\scrs$. Thanks to the regularity of the conformal metric at $\scri^+$ and $i^+$, the estimates both ways between $\scri^+_+$ and $K \cup \scrs$ are straightforward, simply requiring standard Gronwall estimates. We have
\begin{equation} \label{InnerEst}
{\cal E}_{T,\scri^+_+} (\hat\phi )\simeq {\cal E}_{T,{\scrs}} (\hat\phi )+ {\cal E}_{T, K} (\hat\phi )
\end{equation}
the constants in the equivalence depending only on the geometry and not on the solution considered. Outside the domain of influence of $K$, we have estimates both ways between the energies on $\scri^+_- \cup {\scrs}$ and on $\Sigma_0 \setminus K$ using the peeling results~:
\begin{equation} \label{OuterEst}
{\cal E}_{T,\scri^+_-} (\hat\phi )+ {\cal E}_{T,{\scrs}} (\hat\phi ) \simeq {\cal E}_{T, \Sigma_0\setminus K} (\hat\phi ) \, .
\end{equation}
Putting \eqref{InnerEst} and \eqref{OuterEst} together, we obtain the estimates both ways between the energies on $\scri^+$ and $\Sigma_0$~:
\begin{equation} \label{EnEstASST}
{\cal E}_{T,\scri^+} (\hat\phi ) \simeq {\cal E}_{T,\Sigma_0} (\hat\phi ) \, .
\end{equation}
This implies that $T^+$ extends in a unique manner as a linear bounded operator defined on the completion ${\cal H}_{\Sigma_0 }$ of ${\cal C}^\infty_0 (\Sigma_0 ) \times {\cal C}^\infty_0 (\Sigma_0)$ in the norm $\sqrt{ {\cal E}_{T,\Sigma_0} (\hat\phi ) }$ with values in the completion ${\cal H}_{\scri^+ }$ of ${\cal C}^\infty_0 (\scri^+ )$ in the norm $\sqrt{ {\cal E}_{T,\scri^+} (\hat\phi ) }$~; the resulting operator is one-to-one and has closed range. Here ${\cal C}^\infty_0 (\scri^+)$ denotes smooth functions on $\scri^+$ supported away from both $i^+$ and $i_0$. Since $\scri^+$ is of dimension $3$ and $i^+$ is merely a point, assuming the functions supported away from $i^+$ does not impose that the elements of the completion vanish at $i^+$.
\item[Step 3.] In order to show that $T^+$ is onto, we merely need to establish that its range is dense in ${\cal H}_{\scri^+ }$. We do this by solving the Goursat problem from $\scri^+$ for data  $\hat{\phi}_\infty \in {\cal C}^\infty_0 (\scri^+)$. We know from \cite{Ho1990} that \eqref{RescEq} has a unique solution $\hat{\phi} \in {\cal C}^\infty (\hat{\scrm}^+)$ whose restriction to $\scri^+$ is $\hat{\phi}_\infty$. The difficulty is to see that
\begin{equation} \label{ImageGData}
(\hat\phi_0 \, ,~ \hat\phi_1 ) := (\hat\phi \vert_{\Sigma_0} \, , ~ \nabla_T \hat\phi \vert_{\Sigma_0} ) \in {\cal H}_{\Sigma_0 } \, ,
\end{equation}
which will then automatically entail that $\hat{\phi}_\infty = T^+ (\hat\phi_0 \, ,~ \hat\phi_1 )$ as well as the density of the range of $T^+$. The idea is to choose a spacelike hypersurface $\cal S$ in $\scrmhat$, crossing $\scri^+$ in the past of the support of $\hat{\phi}_\infty$ (see figure \ref{IntHyp}). The restriction to $\cal S$ of $\hat\phi$ and $\nabla_T \hat\phi$ are smooth and the crucial observation is that, due to the location of $\cal S$ below the support of the data on $\scri^+$, $\hat\phi \vert_{\cal S}$ vanishes at the boundary of $\cal S$, i.e. at ${\cal S} \cap \scri^+$. The energy norm on $\cal S$ associated with the vector field $T$ is equivalent to the natural $H^1 \times L^2$ norm on $\cal S$ for the rescaled metric $\hat{g}$. Therefore, $\hat\phi \vert_{\cal S} \in H^1_0 ({\cal S})$. It follows that $( \hat\phi \vert_{\cal S} \, ,~ \nabla_T \hat\phi \vert_{\cal S} )$ can be approached, in the energy norm, by a pair of sequences $(\hat\phi_0^n \, ,~ \hat\phi_1^n )$ of smooth functions on $\cal S$, supported away from $\scri^+$. We denote by $\hat\phi^n$ the solution of \eqref{RescEq} in ${\cal C}^\infty (\scrmhat )$ such that
\[ \hat\phi^n \vert_{\cal S} = \hat\phi_0^n \, ,~ \nabla_T \hat\phi^n \vert_{\cal S} = \hat\phi_1^n \, .\]
\begin{figure}[htb!]
\centering
   \includegraphics[width=.65\textwidth]{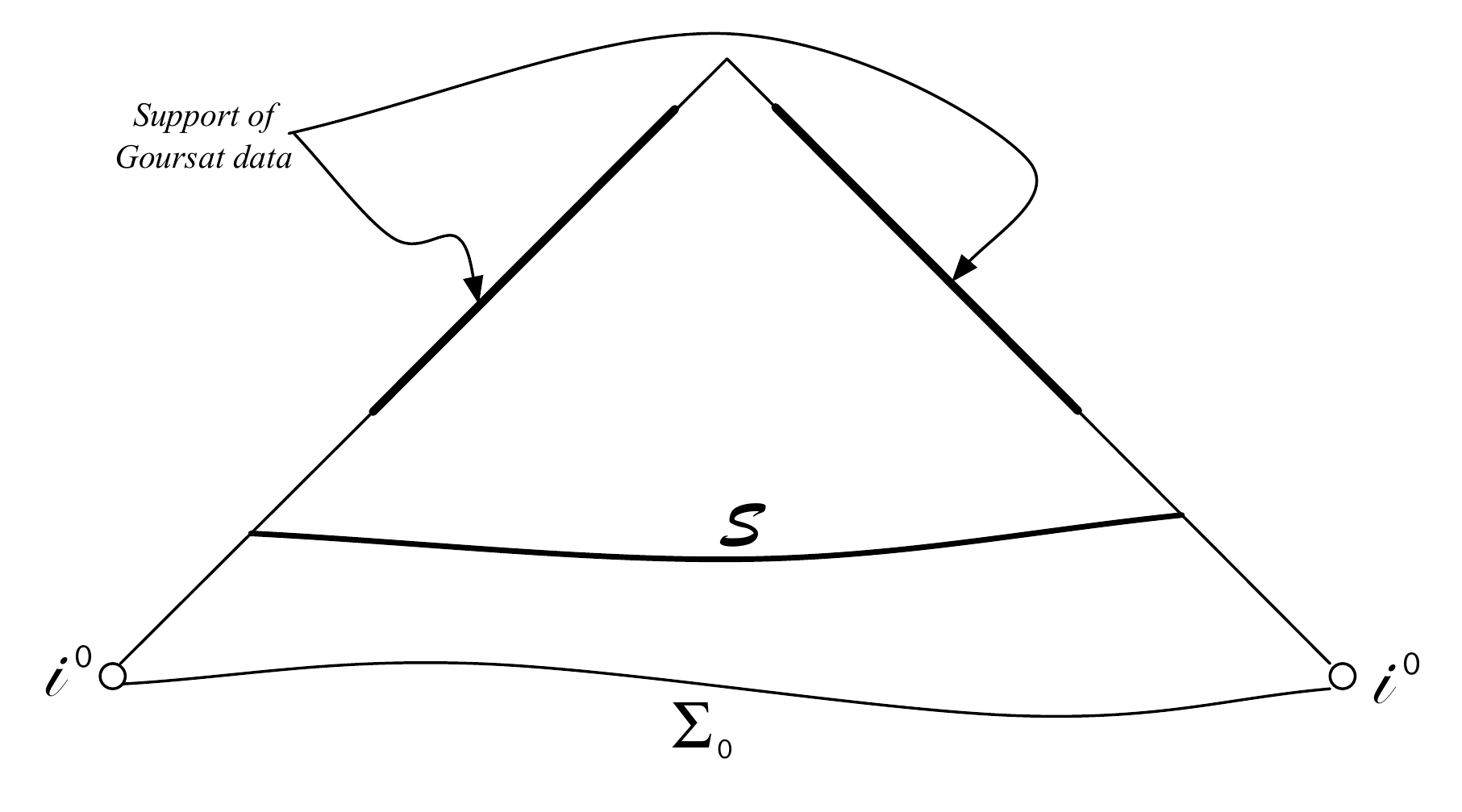}
  \caption{A choice of intermediate hypersurface $\cal S$ for estimating the energy on $\Sigma_0$ of the solution to the Goursat problem.} \label{IntHyp}
\end{figure}
Since by finite propagation speed $\hat\phi^n$ is supported away from $i^0$, we have estimates both ways between the energies of $\hat\phi^n$ on $\cal S$ and on $\Sigma_0$, uniformly in $n$. The convergence of $( \hat\phi^n \vert_{\cal S} \, ,~ \nabla_T \hat\phi^n \vert_{\cal S} )$ towards $( \hat\phi \vert_{\cal S} \, ,~ \nabla_T \hat\phi \vert_{\cal S} )$ in the energy norm on $\cal S$ together with the uniform energy estimates imply that $( \hat\phi^n \vert_{\Sigma_0} \, ,~ \nabla_T \hat\phi^n \vert_{\Sigma_0} )$ is a Cauchy sequence in ${\cal H}_{\Sigma_0}$. This entails that $(\hat\phi_0 \, ,~ \hat\phi_1 ) \in {\cal H}_{\Sigma_0}$.
\end{description}
Therefore the trace operator $T^+$ extends as an isomorphism from ${\cal H}_{\Sigma_0} $ onto ${\cal H}_{\scri^+}$. We can construct $T^-$ in a similar manner and the scattering operator $S = T^+ (T^-)^{-1}$ is then an isomorphism from ${\cal H}_{\scri^- }$ onto ${\cal H}_{\scri^+ }$.

Compared to the case of Minkowski spacetime, the essential change is just the loss of the regularity of the conformal metric at $i^0$, the loss of symmetry is of no importance for our construction. This loss of regularity at $i^0$ is dealt with in a simple manner using essentially the finite propagation speed.

\subsection{Conformal scattering on the Schwarzschild metric} \label{CScatSchwa}

For spacetimes describing isolated black holes in an asymptotically flat universe, the conformal compactification of the exterior is more complicated than for asymptotically simple spacetimes. The singularity at $i^0$ is the same as in the asymptotically simple case, but now the conformal metric is also singular at timelike infinities. This is a much more serious difficulty than the singularity at spacelike infinity. The reason is that finite propagation speed cannot help us here, whatever hypothesis we may make on the supports of the data, $i^+$ will be in their domain of influence. The crucial step is to establish estimates both ways between the energy of the data and that of the trace of the rescaled solution at the conformal boundary. Since the solutions propagate into $i^+$ and the conformal structure is singular there, the regularity of the rescaled field at $i^+$, which allowed to apply Stokes's theorem for the energy current, needs to be replaced by asymptotic information on the behaviour of the field at timelike infinity, typically a sufficiently fast and uniform decay rate in timelike directions. Once this is done, the rest of the construction is mostly unchanged because the behaviour of the conformal metric at $i^+$ has no influence on the propagation of solutions from their future scattering data into the spacetime. We give in this subsection the essential features of the conformal scattering theory developed on the Schwarzschild metric in \cite{Ni2013}. For more technical details as well as a discussion of the additional difficulties in the case of the Kerr metric, see \cite{Ni2013} and references therein, in particular the recent work by M. Dafermos, I. Rodnianski and Y. Shlapentokh-Rothman \cite{DaRoShla}.

\begin{figure}[htb!] 
\centering
\includegraphics[width=4in]{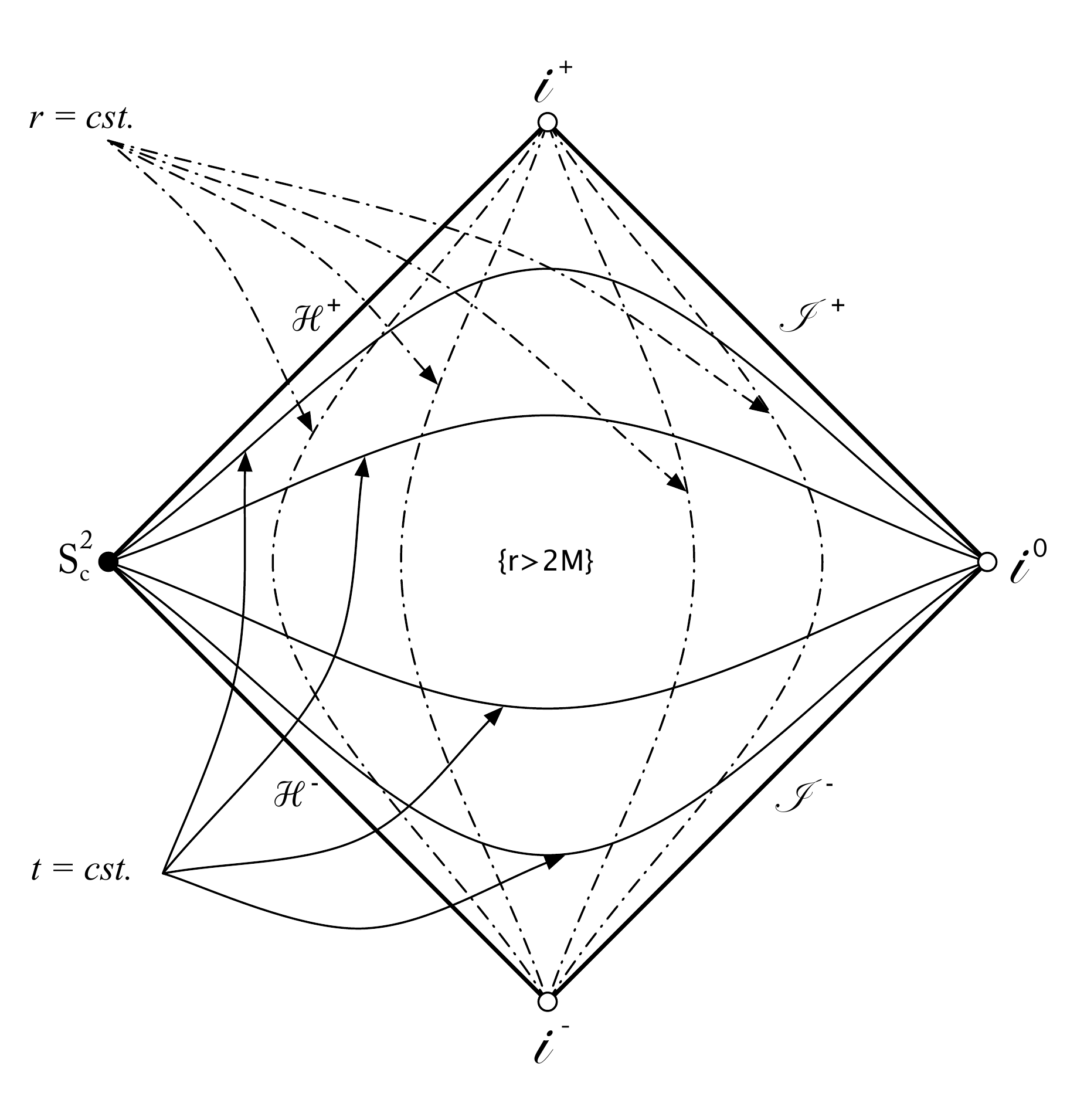}
\caption{The conformal compactification of the exterior of the black hole.} \label{PenD}
\end{figure}

When we considered the conformal compactification of the exterior of the Schwarzschild metric in subsection \ref{SchwaPeel}, we were only interested in constructing $\scri^+$ and in working in the neighbourhood of $i^0$. We are now attempting to develop a conformal scattering theory~; this requires to understand the global geometry of the exterior of the black hole. We still perform the compactification using the simplest conformal factor $\Omega = 1/r$, but we look at the rescaled metric using different coordinate systems in order to construct all the components of the boundary. Recall that the Schwarzschild metric is given on $\R_t \times ]0,+\infty [_r \times S^2_\omega$ by
\[ g = F \d t^2 - F^{-1} \d r^2 - r^2 \d \omega^2 \, ,~ F = 1-2m/r \]
and the change of radial variable $r_* = r + 2m \log (r-2m )$ maps the exterior of the black hole $\R_t \times ]2m , +\infty [_r \times S^2_\omega$ to the domain $\R_t \times \R_{r_*} \times S^2_\omega$ with the new expression for $g$
\[ g = F (\d t^2 - \d r_*^2 ) - r^2 \d \omega^2 \, .\]
Using coordinates $u = t-r_*$, $R=1/r$, $\omega$, the rescaled metric $\hat{g} = R^2 g$ reads
\[
\hat{g} = R^2 (1-2mR) \d u^2 - 2 \d u \d R - \d \omega^2 \, .
\]
In this coordinate system, $\scri^+$ and the past horizon $\scrh^-$ appear as the smooth null hypersurfaces
\[ \scri^+ = \R_u \times \{ 0\}_R \times S^2_\omega \, ,~ \scrh^- = \R_u \times \{ 1/2m \}_R \times S^2_\omega \, .\]
Similarly, using the coordinates $v=t+r_*,R,\omega$, the metric $\hat{g}$ takes the form
\[
\hat{g} = R^2 (1-2mR) \d v^2 + 2 \d v \d R - \d \omega^2 \, .
\]
We now have access to past null infinity $\scri^-$ and to the future horizon $\scrh^+$, appearing respectively as the smooth null hypersurfaces
\[ \scri^- = \R_v \times \{ 0\}_R \times S^2_\omega \, ,~ \scrh^+ = \R_v \times \{ 1/2m \}_R \times S^2_\omega \, .\]
At the past and future horizons, not only the rescaled metric, but also the physical metric $g$, extends analytically as a non degenerate metric~; $\scrh^+$ and $\scrh^-$ meet at a $2$-sphere $S^2_c$, called the crossing sphere, at which both $g$ and $\hat{g}$ extend analytically and are non degenerate. The construction of $S^2_c$ can be done using Kruskal-Szekeres coordinates (see for example \cite{HaEl} or \cite{Wa}). The Penrose diagram of the compactified exterior is given in figure \ref{PenD}. Instead of two null hypersurfaces diffeomorphic to $\R \times S^2$, the boundary of our compactified spacetime now contains four such hypersurfaces.
\begin{remark}
Note that the different components of the boundary of the rescaled exterior of the black hole are of two quite different natures. Null infinities $\scri^\pm$ on the one hand are genuinely part of the conformal boundary of the spacetime, describing ``points at infinity''. The horizons $\scrh^\pm$ on the other hand do not describe points at infinity for $g$ but the finite boundary of the exterior of the black hole. They are part of the boundary of our compactified spacetime only because we restrict our study to the exterior of the black hole. This is justified by the fact that our scattering theory is assumed to reflect the point of view of an observer static at infinity, whose perception does not go beyond the horizon.
\end{remark}
We consider the Cauchy hypersurface
\[ \Sigma_0 = \{ 0 \}_t \times \R_{r_*} \times S^2_\omega \, .\]
In order to establish estimates both ways between $\Sigma_0$ and $\scri^+ \cup \scrh^+$, we proceed in two steps.
\begin{enumerate}
\item For $T>0$, we consider the three hypersurfaces\footnote{We give here an explicit choice of hypersurface $S_T$, but the only important properties that $S_T$ needs to satisfy are that it is achronal for the rescaled metric and that $\Sigma_0\cup \scrh^+_T \cup S_T \cup \scri^+_T$ forms a closed hypersurface (except where $\scri^+$ and $\Sigma_0$ meet $i^0$).} (see Figure \ref{3surface})
\begin{figure}[ht] 
\centering
\includegraphics[width=4in]{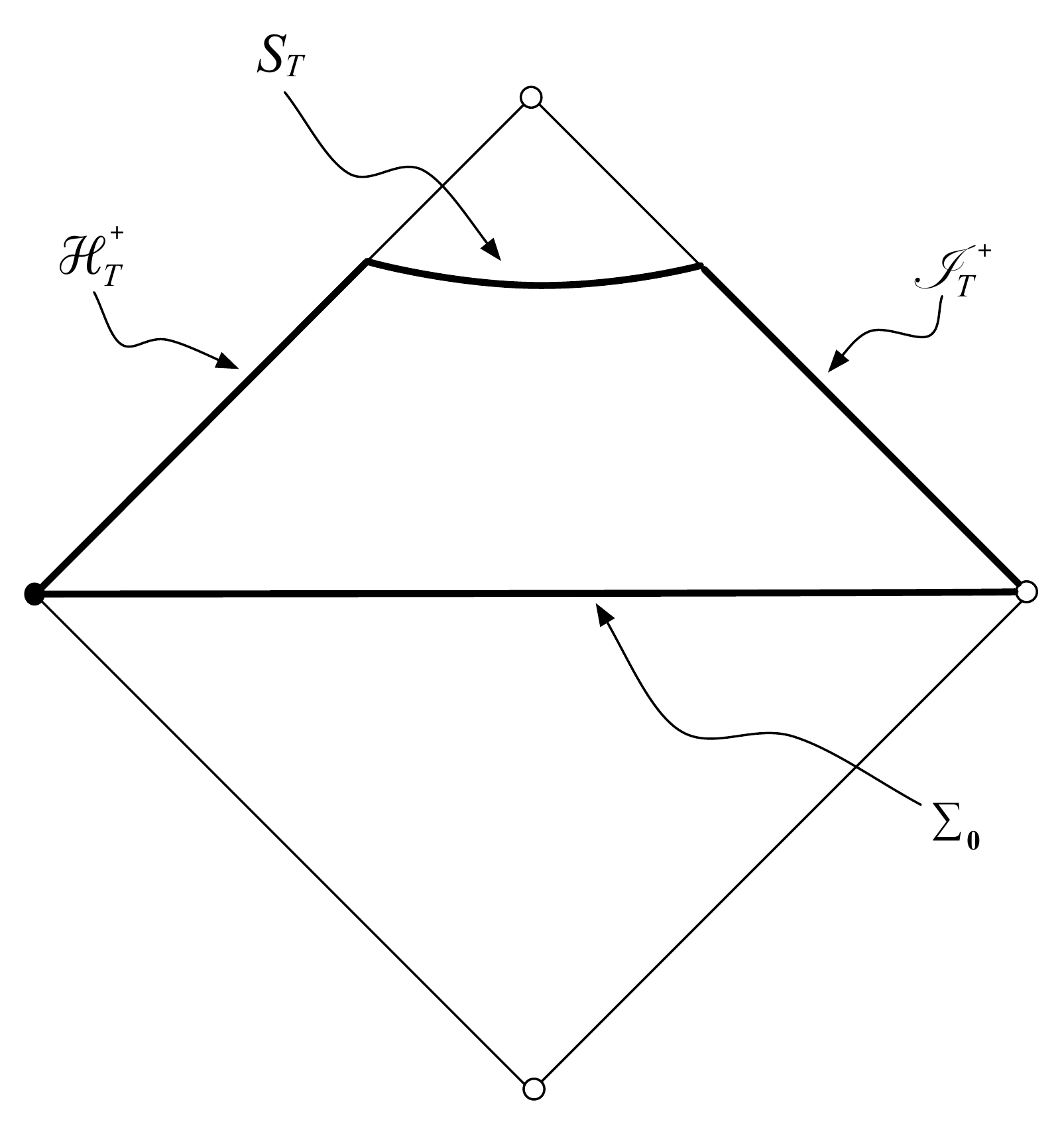}
\caption{The main hypersurfaces represented on the compactified exterior.}  \label{3surface}
\end{figure}
\begin{eqnarray}
S_T &=& \left\{ (t,r_* , \omega) \in \R \times \R \times S^2 \, ;~ t = T+ \sqrt{1+r_*^2} \right\} \label{ST} \, , \\
\scri^+_T &=& \scri^+ \cap \{ u \leq T\} = ]-\infty , T]_u \times \{ 0 \}_R \times S^2_\omega \, , \label{scriT} \\
\scrh^+_T &=& S^2_{\mathrm{c}} \cup (\scrh^+ \cap \{ v \leq T\} ) = S^2_{\mathrm{c}} \cup (]-\infty , T]_v \times \{ 1/2m \}_R \times S^2_\omega ) \, . \label{scrhT}
\end{eqnarray}
We establish energy estimates both ways between $\Sigma_0$ and $\scrh^+_T \cup S_T \cup \scri^+_T$, uniformly in $T>0$. We use the stress-energy tensor \eqref{SETCWESchwa} for the wave equation \eqref{SchwaConfWaveEq} associated with $\hat{g}$. We recall its conservation law \eqref{ConsLawSETCWESchwa} for $\hat\phi$ solution of \eqref{SchwaConfWaveEq}~:
\[ \hat\nabla^a \hat{T}_{ab} = -2mR \hat\phi \hat\nabla_b \hat\phi \, .\]
This entails that the energy current
\[ \hat{J}_a = K^b \hat{T}_{ab} \, ,~ K = \partial_t \, ,\]
is not conserved but satisfies instead
\[ \hat\nabla^a \hat{J}_a = -2mR \hat\phi \partial_t \hat\phi \, . \]
However, thanks to the symmetries of Schwarzschild's spacetime, this equation can easily be seen as the exact conservation law
\begin{equation} \label{ConsLawSchwa}
\hat\nabla_a \left( \hat{J}^a + V^a \right) =0 \, ,~ \mbox{with } V = mR\hat\phi^2 \partial_t \, .
\end{equation}
Since $V$ is causal and future oriented, we still have that the flux of the modified current $J+V$ defines a positive definite (resp. non negative) quadratic form on spacelike (resp. achronal) hypersurfaces. We introduce the modified energy on an oriented hypersurface $S$
\[
\hat{\cal E}_{\partial_t, S} = \int_{S} (\hat{J}_a + V_a )n^a ( l \lrcorner \dvol ) \, ,\]
where $l^a$ is a vector field transverse to $S$ compatible with the orientation of $S$ and $n^a$ a normal vector field to $S$ such that $\hat{g} (l,n) =1$. For any $T>0$, we have
\begin{equation} \label{EnIdentityT}
\hat{\cal E}_{\partial_t, \Sigma_0} = \hat{\cal E}_{\partial_t, \scri^+_T} + \hat{\cal E}_{\partial_t, \scrh^+_T} + \hat{\cal E}_{\partial_t, S_T} \, .
\end{equation}
\item We take $T$ to $+\infty$. The modified energies on $\scrh^+_T$ and $\scri^+_T$ tend to $\hat{\cal E}_{\partial_t, \scrh^+}$ and $\hat{\cal E}_{\partial_t, \scri^+}$ respectively. The last thing we need in order to conclude is to show that $\hat{\cal E}_{\partial_t, S_T}$ tends to zero. On the Schwarzschild metric, we know enough on the decay of solutions to the wave equation to infer this (see M. Dafermos and I. Rodnianski \cite{DaRo}). Generally for this type of approach, the expected generic decay known as Price's law will be sufficient to establish that $\hat{\cal E}_{\partial_t, S_T}$ tends to zero.
\end{enumerate}
As mentioned above, the rest of the construction is essentially unchanged.

\section{Concluding remarks} \label{Concl}

We have described two approaches to asymptotic analysis making a fundamental use of conformal compactifications. Both constructions in fact rely on a choice of spacelike hypersurface as an intermediate tool. In the case of conformal scattering, the Cauchy hypersurface is used to construct the trace operators $T^\pm$, which play the role of inverse wave operators. The object that the theory aims to construct is the scattering operator, mapping the past radiation field to the future radiation field. This operator is independent of the choice of spacelike hypersurface from which the trace operators are defined and the theory is in fact truly covariant. For our approach of the peeling, the choice of Cauchy hypersurface is more fundamental since we study the asymptotic properties of Cauchy data that entail a certain transverse regularity of the rescaled solution at null infinity. However, provided we only work with asymptotically flat Cauchy hypersurfaces, these asymptotic properties ought to be independent of the choice of Cauchy hypersurface and in this sense the theory could also be understood as covariant. An interesting alternative (and much more delicate) approach to the peeling would be to characterize the transverse regularity at $\scri^+$ in terms of the function space of past scattering data.

\begin{center}
{\bf Acknowledgments}
\end{center}

The author would like to thank the organizer of the conference, Athanase Papadopoulos, for the invitation. The research presented in this essay was partly supported by the ANR funding ANR-12-BS01-012-01.


\begin{thebibliography}{100}
\bibitem{Ba1989a} J.C. Baez, {\em Scattering and the geometry of the solution manifold of $\square f+ \lambda f^3 =0$}, J. Funct. Anal. {\bf 83} (1989), 317--332.

\bibitem{Ba1989b} J.C. Baez, {\em Scattering for the Yang-Mills equations},  Trans. Amer. Math. Soc. {\bf 315} (1989), 2, 823--832.

\bibitem{Ba1990} J.C. Baez, {\em Conserved quantities for the Yang-Mills equations}, Adv. Math. {\bf 82} (1990), 1, 126--131.

\bibitem{BaSeZho1990}  J.C. Baez, I.E. Segal, Zhou Z.F., {\em The global Goursat problem and scattering for nonlinear wave equations}, J. Funct. Anal. {\bf 93} (1990), 2, 239--269.

\bibitem{BaZho1989} J.C. Baez, Zhou Z.F., {\em The global Goursat problem on $\R \times S^1$}, J. Funct. Anal. {\bf 83} (1989), 364--382.

\bibitem{ChruDe2002} P. Chrusciel \& E. Delay, {\em Existence of non trivial,
    asymptotically vacuum, asymptotically simple space-times},
    Class. Quantum Grav. {\bf 19} (2002), L71-L79, erratum
    Class. Quantum Grav. {\bf 19} (2002), 3389.

\bibitem{ChruDe2003} P. Chrusciel \& E. Delay, {\em On mapping properties of
    the general relativistic constraints operator in weighted function
    spaces, with applications}, Mém. Soc. Math. Fr. (N.S.) {\bf 94} (2003), vi+103 pp.

\bibitem{Co2000} J. Corvino, {\em Scalar curvature deformation and a
    gluing construction for the Einstein constraint equations},
    Comm. Math. Phys. {\bf 214} (2000), 137--189.

\bibitem{CoScho2003} J. Corvino \& R.M. Schoen, {\em On the asymptotics
    for the vacuum Einstein constraint equations}, gr-qc 0301071 (2003), J. Differential Geom. {\bf 73} (2006), no. 2, 185--217.

\bibitem{DaRo} M. Dafermos \& I. Rodnianski, {\em The redshift effect and radiation decay on black hole space-times}, Comm. Pure Appl. Math. {\bf 62} (2009), 7, 859-919.

\bibitem{DaRoShla} M. Dafermos, I. Rodnianski, Y. Shlapentokh-Rothman, {\em A scattering theory for the wave equation on Kerr black hole exteriors}, arXiv:1412.8379.

\bibitem{Di1985} J. Dimock, {\em Scattering for the wave equation on the Schwarzschild metric}, Gen. Rel. Grav. {\bf 17} (1985), 4, 353--369.

\bibitem{DiKa1986} J. Dimock, B.S. Kay, {\em Classical and quantum scattering theory for linear scalar fields on the Schwarzschild metric. II}, J. Math. Phys. {\bf 27} (1986), 10, 2520--2525.

\bibitem{DiKa1987} J. Dimock, B.S. Kay, {\em Classical and Quantum Scattering theory for linear scalar fields on the Schwarzschild metric I}, Ann. Phys. {\bf 175} (1987),  366--426.

\bibitem{Fri1962} F.G. Friedlander, {\em On the radiation field of pulse solutions of the wave equation}, Proc. Roy. Soc. Ser. A {\bf 269} (1962), 53--65.

\bibitem{Fri1964} F.G. Friedlander, {\em On the radiation field of pulse solutions of the wave equation II}, Proc. Roy. Soc. Ser. A {\bf 279} (1964), 386--394.

\bibitem{Fri1967} F.G. Friedlander, {\em On the radiation field of pulse solutions of the wave equation III}, Proc. Roy. Soc. Ser. A {\bf 299} (1967), 264--278.

\bibitem{Fri1980} F.G. Friedlander, {\em Radiation fields and hyperbolic scattering theory}, Math. Proc. Camb. Phil. Soc. {\bf 88} (1980), 483--515.

\bibitem{Fri2001} F.G. Friedlander, {\em Notes on the Wave Equation on Asymptotically Euclidean Manifolds}, J. Funct. Anal. {\bf 184} (2001), 1--18.

\bibitem{HaEl} S.W. Hawking \& G.F.R Ellis, {\em The large scale structure of space-time}, Cambridge monographs in mathematical physics, Cambridge University Press 1973.

\bibitem{Ho1990} L. Hörmander, {\em A remark on the characteristic Cauchy problem}, J. Funct. Ana. {\bf 93} (1990), 270--277.

\bibitem{InglNi} W. Inglese, F. Nicolò, {\em Asymptotic properties of the electromagnetic field in the external Schwarzschild spacetime}, Ann. Henri Poincaré {\bf 1} (2000), 5, 895-944.

\bibitem{Jo2012} J. Joudioux, {\em Conformal scattering for a nonlinear wave equation}, J. Hyperbolic Differ. Equ. {\em 9} (2012), 1, 1--65.

\bibitem{LaPhi} P.D. Lax, R.S. Phillips, {\em Scattering theory}, Academic Press 1967.

\bibitem{Le1953} J. Leray, {\em Hyperbolic differential equations}, lecture notes, Princeton Institute for Advanced Studies, 1953.

\bibitem{MaNi2004} L.J. Mason, J.-P. Nicolas, {\em Conformal scattering and the Goursat problem}, J. Hyperbolic Differ. Equ. {\bf 1} (2004), 2, 197--233.

\bibitem{MaNi2009} L.J. Mason, J.-P. Nicolas, {\em Regularity at space-like and null infinity}, J. Inst. Math. Jussieu {\bf 8} (2009), 1, 179--208.

\bibitem{MaNi2012} L.J. Mason, J.-P. Nicolas, {\em Peeling of Dirac and Maxwell fields on a Schwarzschild background}, J. Geom. Phys. {\bf 62} (2012), 867--889.

\bibitem{Mo1961} C.S. Morawetz, {\em The decay of solutions of the exterior initial-boundary value problem for the wave equation}, Comm. Pure Appl. Math. {\bf 14} (1961), p. 561--568.

\bibitem{NP} E.T. Newman, R. Penrose, {\em An approach to gravitational radiation by a method of spin coefficients}, J. Math. Phys. {\bf 3} (1962), 566--768.

\bibitem{Ni2013} J.-P. Nicolas, {\em Conformal scattering on the Schwarzschild metric}, arXiv:1312.1386.

\bibitem{Pe1963} R. Penrose, {\em Asymptotic properties of fields and space-times}, Phys. Rev. Lett. {\bf 10} (1963), 66--68.

\bibitem{Pe1964} R. Penrose, {\em Conformal treatment of infinity}, 1964 Relativité, Groupes et Topologie (Lectures, Les Houches, 1963 Summer School of Theoret. Phys., Univ. Grenoble) 563--584 Gordon and Breach, New York.

\bibitem{Pe1965} R. Penrose, {\em Zero rest-mass fields including gravitation~: asymptotic behaviour}, Proc. Roy. Soc. London {\bf A284} (1965), 159--203.

\bibitem{PeRi1984} R. Penrose and W. Rindler, Spinors and space-time, Vol. I (1984) and Vol. 2 (1986), Cambridge University Press.

\bibitem{Sa61} R. Sachs, {\em Gravitational waves in general relativity VI, the outgoing radiation condition}, Proc. Roy. Soc. London {\bf A264} (1961), 309-338.

\bibitem{Sa62} R. Sachs, {\em Gravitational waves in general relativity VIII, waves in asymptotically flat space-time}, Proc. Roy. Soc. London {\bf A270} (1962), 103--126.

\bibitem{Wa} R. Wald, {\em General Relativity}, University of Chicago Press, 1984.

\bibitem{Whi1903} E.T. Whittaker, {\em On the partial differential equations of mathematical physics}, Mathematische Annalen {\bf 57} (1903), p. 333-355.
\end{thebibliography}
\end{document}